\documentclass[12pt]{article}
\usepackage{axodraw,epsfig,cite}

\input paperdef

\evensidemargin \oddsidemargin
\def\_{\rule{.3em}{.15ex}} 

\newcommand{\scs}{\scriptscriptstyle}

\def\slash#1{\setbox0=\hbox{$#1$}#1\hskip-\wd0\dimen0=5pt\advance
       \dimen0 by-\ht0\advance\dimen0 by\dp0\lower0.5\dimen0\hbox
         to\wd0{\hss\sl/\/\hss}}
\def\simleq{\stackrel{<}{\scs \sim}}

\newcommand{\MHaa}{{\cal M}_H\Big|_{aa}}

\begin{document}
\thispagestyle{empty}

\def\thefootnote{\fnsymbol{footnote}}

\begin{flushright}
BNL--HET--01/27\\
hep-ph/0108059
\end{flushright}

\vspace{1cm}

\begin{center}

{\large\sc {\bf The Higgs Boson Sector of the Complex MSSM}}

\vspace{0.4cm}

{\large\sc {\bf in the Feynman-diagrammatic approach}}
 
\vspace{1cm}

{\sc 
S.~Heinemeyer$^{1}$%
\footnote{email: Sven.Heinemeyer@bnl.gov}%
}

\vspace*{1cm}

{\sl
$^1$ HET, Brookhaven Natl.\ Lab., Upton, New York 11973, USA
}

\end{center}

\vspace*{0.2cm}

\begin{abstract}
  In the Minimal Supersymmetric Standard Model with complex parameters\\
  (cMSSM) we calculate higher order corrections to the Higgs boson
  sector in the Feynman-diagrammatic approach using the on-shell
  renormalization scheme. The application of this
  approach to the cMSSM, being complementary to existing approaches,
  is analyzed in detail. Numerical examples for
  the leading fermionic corrections, including the leading \twol\
  effects, are presented. Numerical agreement within 10\% with other
  approaches is found for small and moderate mixing in the scalar top
  sector. The leading fermionic 
  corrections, supplemented by the full logarithmic \onel\ and the
  leading \twol\ 
  contributions are implemented into the public Fortran code \fhfc.
\end{abstract}

\def\thefootnote{\arabic{footnote}}
\setcounter{page}{0}
\setcounter{footnote}{0}

\newpage


\section{Introduction}

The search for the lightest Higgs boson is a crucial test of 
Supersymmetry (SUSY) which can be performed with the present and the
next generation of accelerators. The prediction of a relatively
light Higgs boson is common to all supersymmetric models whose
couplings remain in the perturbative regime up to a very high energy
scale~\cite{susylighthiggs}.
A precise prediction for the mass of the lightest Higgs boson and its
couplings to other particles in terms
of the relevant SUSY parameters is necessary in order to determine the
discovery and exclusion potential of LEP2 and the upgraded Tevatron, and
for physics at the LHC and future linear colliders, where eventually a
high-precision measurement of the properties of the Higgs boson might
be possible~\cite{teslatdr}. 

The case of the Higgs sector in the $\cp$-conserving MSSM has been
tackled up to the \twol\ level by different methods such as the
Effective Potential (EP) method~\cite{mhiggsEP}, the renormalization
group (RG) improved \onel\ EP approach~\cite{mhiggsRG} and the
Feynman-diagrammatic (FD) method using the on-shell renormalization
scheme~\cite{mhiggsletter,mhiggslong}. The application of different
methods lead to thorough comparisons between the different approaches.
Most prominently the comparison between the RG improved
\onel\ EP 
result and the FD result~\cite{bse,mhiggsRGFD2,mhiggsRGFD3}, and most recently
between the FD 
and the EP result~\cite{mhiggsRGFD3,mhiggsEPFD}, have been performed. 
These comparisons, showing 
agreement where expected, lead to deeper insight into the radiative
corrections in the MSSM Higgs sector and thus to the confidence that
the higher-order contribution, although being large, are under
control.

In the case of the MSSM with complex parameters (cMSSM) the higher
order corrections have yet been restricted, after the first more
general investigations~\cite{mhiggsCPXgen}, to evaluations in the EP
approach~\cite{mhiggsCPXEP1,mhiggsCPXEP2} and to the RG improved \onel\ EP
method~\cite{mhiggsCPXRG1,mhiggsCPXRG2}. While in the MSSM without
complex parameters the FD calculation, using the on-shell
renormalization scheme, has provided the only complete calculation at
the \onel\ level~\cite{mhiggsFD1l} and furthermore the relevant
logarithmic and non-logarithmic 
corrections at the \twol\ level~\cite{mhiggslong,bse}, a corresponding
calculation in the cMSSM has been missing so far.

This paper provides the next step into this direction: it is shown in
detail how the FD method, employing the on-shell renormalization
scheme, can be applied to the Higgs sector of the cMSSM. The general
analysis is exemplified at the leading fermionic \onel\ corrections,
showing the applicability of the method and providing the full
corresponding analytical result. For numerical examples and the
comparison with existing approaches, the result is supplemented by
non-leading corrections at the one- and \twol\ level taken over from
the real MSSM case. All results are finally incorporated into a public
Fortran code. 
A detailed analysis, including a full \onel\ calculation and the
dominant \twol\ corrections to the cMSSM Higgs sector will be
presented elsewhere~\cite{mhiggsCPXFD}.

The rest of the paper is organized as follows. In Section~2 we review
the Higgs sector and the scalar quark sector of the cMSSM, providing
all relevant information about the relations of physical and
unphysical parameters, the masses and the mixing
angles. The renormalization in the on-shell scheme in the cMSSM Higgs
sector is presented in detail in Section~3, together with the
analytical result for the leading fermionic corrections obtained in
this approach. Section~4 briefly reviews the evaluation of the Higgs
boson masses and couplings in the FD approach. Numerical results for
the comparison with other approaches are given in Section~5. Section~6
contains the description of the corresponding Fortran code \fhfc. The
conclusions can be found in Section~7.


\section{Calculational basis}

\subsection{The tree-level Higgs sector of the cMSSM}

The (c)MSSM Higgs potential reads~\cite{hhg}:
\BEA
\label{Higgspot}
V &=& m_1^2 \cHe\bar{\cHe} + m_2^2 \cHz\bar{\cHz} - m_{12}^2 (\epsilon_{ab}
      \cHe^a\cHz^b + \hc)  \nonumber \\
   && \mbox{} + \frac{g'^2 + g^2}{8}\, (\cHe\bar{\cHe} - \cHz\bar{\cHz})^2
      +\frac{g^2}{2}\, |\cHe\bar{\cHz}|^2,
\EEA
where $m_1^2, m_2^2, m_{12}^2$ are soft SUSY-breaking terms, $g, g'$
are the $SU(2)$ and $U(1)$ gauge couplings, and $\epsilon_{12} = -1$.
The doublet fields $\cHe$ and $\cHz$ are decomposed  in the following way:
\BEA
\cHe &=& \VL H_1^1 \\ H_1^2 \VR = \VL v_1 + (\phi_1^{0} + i\chi_1^{0})
                                 /\sqrt2 \\ \phi_1^- \VR ,\non \\
\cHz &=& \VL H_2^1 \\ H_2^2 \VR =  e^{i\xi} \VL \phi_2^+ \\ v_2 + (\phi_2^0 
                                              + i\chi_2^0)/\sqrt2 \VR.
\label{eq:hidoubl}
\EEA
$\xi$ is a possible new phase between the two Higgs doublets. From the
unphysical parameters in \refeq{Higgspot} the transition to the
physical parameters (including the tadpoles) is performed by the
following substitution (see 
also \citeres{mhiggslong,mhiggsCPXgen}):
\BEA
v_1 &\to& \frac{\wz \cbe \sw \cw \MZ}{e} \non \\
v_2 &\to& \frac{\wz \sbe \sw \cw \MZ}{e} \non \\
g_1 &\to& \frac{e}{\cw} \non \\
g_2 &\to& \frac{e}{\sw} \non \\
m_1^2 &\to& \MHpq^2 \sbe^2 - \edz (\cbe^2 - \sbe^2) \MZ^2
            + t_1 \frac{e}{2 \sw \cw \MZ} \cbe (1 + \sbe^2) \non \\
 &&         - t_2 \frac{e}{2 \sw \cw \MZ} \sbe \cbe^2 \non \\
m_2^2 &\to& \MHpq^2 \cbe^2 + \edz (\cbe^2 - \sbe^2) \MZ^2
            - t_1 \frac{e}{2 \sw \cw \MZ} \cbe \sbe^2 \non \\
 &&         + t_2 \frac{e}{2 \sw \cw \MZ} \sbe (1 + \cbe^2) \non \\
\re m_{12}^2 &\to& \KL - \MHpq^2 \sbe \cbe
                       + t_1 \frac{e}{2 \sw \cw \MZ} \sbe^3
                       + t_2 \frac{e}{2 \sw \cw \MZ} \cbe^3 \KR 
                    \ed{\cos\xi} \non\\
\im m_{12}^2 &\to& \KL t_A \frac{e}{2 \sw \cw \MZ} \KR
                    \ed{\sin\xi} .
\EEA
$\tb$ is the ratio of the two vacuum expectation values, $\tb =
v_2/v_1$, and $\sbe = \Sb, \cbe = \Cb$, $\cw \equiv \MW/\MZ$,
$\sw^2 = 1 - \cw^2$. 
$\MHpq^2 \equiv \MHp^2 - \MW^2$, where (as will be shown
below) $\MHp$ is the mass of the
charged Higgs boson $H^\pm$.
Contrary to the real case, where the mass of the $\cp$-odd
Higgs boson, $\MA$, is used as input parameter, in
the cMSSM $\MHp$ is chosen as physical parameter,
since the field $A \equiv \sbe \Ce + \cbe \Cz$ 
(as will be shown later) mixes with the fields
$\Pe$ and $\Pz$.
$t_1$ and $t_2$ denote the tadpoles of the fields $\Pe$ and
$\Pz$, whereas $t_A$ is the tadpole of the field $A$.
The expressions for the tadpoles can be obtained directly by expanding
the Higgs potential \refeq{Higgspot} in the fields from the terms
linear in $\Pe, \Pz$ and $A$.

In the cMSSM all neutral Higgs bosons can
mix. Therefore the following $(4 \times 4)$ mass matrix has to be
considered~\cite{mhiggsCPXgen}:
\BE
M_{\rm Higgs} = \ML M_\rmS  & M_\rmSP \\
                      M^+_\rmSP & M_\rmP \MR ,
\EE
resulting in the Lagrange density
\BE
\cL = \edz\; (\Pe, \Pz, \Ce, \Cz) \quad M_\rHiggs \quad
           \VL \Pe \\ \Pz \\ \Ce \\ \Cz \VR .
\EE
Here $M_\rmS$ denotes the $(2 \times 2)$ mass matrix of the fields
$\Pe$ and $\Pz$ (the $\cp$-even mass matrix in the real MSSM), 
$M_\rmP$ represents the $(2 \times 2)$ mass matrix of the fields
$\Ce$ and $\Cz$ (the $\cp$-odd mass matrix in the real MSSM).
$M_\rmSP$ denotes the mixing terms (which are always zero in the
real MSSM). The three matrices are given in terms of physical
parameters by 
\BEA
M_\rmS &=& \ML \mpe^2 & \mpez^2 \\ \mpez^2 & \mpz^2 \MR \\
 &=&        \ML \MHpq^2 \sbe^2 + \MZ^2 \cbe^2 &
               -\sbe \cbe (\MHpq^2 + \MZ^2) \\
               -\sbe \cbe (\MHpq^2 + \MZ^2) &
               \MHpq^2 \cbe^2 + \MZ^2 \sbe^2 \MR \non \\
 && + \ML           \bar t_1  \cbe (1 + \sbe^2)
                  - \bar t_2  \sbe \cbe^2 &
                    \bar t_1  \sbe^3
                  + \bar t_2  \cbe^3 \\
                    \bar t_1  \sbe^3
                  + \bar t_2  \cbe^3  &
                  - \bar t_1  \cbe \sbe^2
                  + \bar t_2  \sbe (1 + \cbe^2) \MR \\[1em]
M_\rmSP &=& \ML 0 &
                \bar t_A \\
                \bar t_A &
                0 \MR \\[1em]
M_\rmP &=& \ML \MHpq^2 \sbe^2 &
               \MHpq^2 \sbe \cbe \\
               \MHpq^2 \sbe \cbe &
               \MHpq^2 \cbe^2 \MR \non \\
 && + \ML           \bar t_1  \cbe (1 + \sbe^2)
                  - \bar t_2  \sbe \cbe^2 &
                  - \bar t_1  \sbe^3
                  - \bar t_2  \cbe^3 \\
                  - \bar t_1  \sbe^3
                  - \bar t_2  \cbe^3 &
                  - \bar t_1  \cbe \sbe^2
                  + \bar t_2  \sbe (1 + \cbe^2) \MR
\EEA
with $\bar t_x \equiv t_x \, e/(2 \sw \MW), \; x = 1, 2, A$.\\
Similarly the matrix of the charged Higgs bosons is given by
\BEA
M_\rmC &=& \ML \MHp^2 \sbe^2 &
               \MHp^2 \sbe \cbe \\
               \MHp^2 \sbe \cbe &
               \MHp^2 \cbe^2 \MR \non \\
 && + \ML           \bar t_1  \cbe (1 + \sbe^2)
                  - \bar t_2  \sbe \cbe^2 &
                  - \bar t_1  \sbe^3
                  - \bar t_2  \cbe^3 \\
                  - \bar t_1  \sbe^3
                  - \bar t_2  \cbe^3 &
                  - \bar t_1  \cbe \sbe^2
                  + \bar t_2  \sbe (1 + \cbe^2) \MR
\EEA


\subsection{Rotation with $\be$}

The angle $\be$ diagonalizes (up to tadpole contributions) the matrix
$M_\rmP$: 
\BE
\VL G \\ A \VR = D^+(\be) \VL \Ce \\ \Cz \VR
               =  \ML \cbe & -\sbe \\ \sbe & \cbe \MR \VL \Ce \\ \Cz \VR
\EE
\BEA
&& (\Ce, \Cz) \; M_\rmP \; \VL \Ce \\ \Cz \VR \non \\
&=& (\Ce, \Cz) \; D(\be)D^+(\be) \; M_\rmP \; D(\be)D^+(\be) 
   \VL \Ce \\ \Cz \VR \non \\
&=& (G, A) \; M^D_\rmP \; \VL G \\ A \VR
\EEA
with 
\BE
M^D_\rmP = \ML \cbe \bar t_1 + \sbe \bar t_2 &
               \sbe \bar t_1 - \cbe \bar t_2 \\
               \sbe \bar t_1 - \cbe \bar t_2 &
               \MHpq^2 \MR .
\EE
This also affects the matrix $M_\rmSP$. Defining the $(4 \times 4)$
matrix 
\BE
D^4(\be) = \ML \id & 0 \\ 0 & D(\be) \MR ,
\EE
the rotation of $M_\rHiggs$ can be performed:
\BEA
&&  (\Pe, \Pz, \Ce, \Cz) \; M_\rHiggs \; 
    \VL \Pe \\ \Pz \\ \Ce \\ \Cz \VR \non \\
&=&  (\Pe, \Pz, \Ce, \Cz) \; D^4(\be)D^{4+}(\be) \; M_\rHiggs \; 
    D^4(\be)D^{4+}(\be) \; \VL \Pe \\ \Pz \\ \Ce \\ \Cz \VR \non \\
&=& (\Pe, \Pz, G, A) \; M^\be_\rHiggs \; 
    \VL \Pe \\ \Pz \\ G \\ A \VR
\EEA
with
\BE
M^\be_\rHiggs = \ML M_\rmS  & M^\be_\rmSP \\
                      M^{\be+}_\rmSP & M^D_\rmP \MR
\EE
and
\BE
M^\be_\rmSP = \bar t_A \ML -\sbe  & \cbe  \\
                            \cbe  & \sbe  \MR .
\EE

The angle $\be$ diagonalizes (up to tadpole contributions) also the matrix
$M_\rmC$: 
\BE
\VL G^\pm \\ H^\pm \VR = D^+(\be) \VL \Pe^\pm \\ \Pz^\pm \VR
    =  \ML \cbe & -\sbe \\ \sbe & \cbe \MR \VL \Pe^\pm \\ \Pz^\pm \VR
\EE
\BEA
&& (\Pe^-, \Pz^-) \; M_\rmC \; \VL \Pe^+ \\ \Pz^+ \VR \non \\
&=& (\Pe^-, \Pz^-) \; D(\be)D^+(\be) \; M_\rmC \; D(\be)D^+(\be) 
   \VL \Pe^+ \\ \Pz^+ \VR \non \\
&=& (G^-, H^-) \; M^D_\rmC \; \VL G^+ \\ H^+ \VR
\EEA
with 
\BE
M^D_\rmC = \ML \cbe \bar t_1 + \sbe \bar t_2 &
               +\sbe \bar t_1 - \cbe \bar t_2 \\
               +\sbe \bar t_1 - \cbe \bar t_2 &
               \MHp^2  \MR .
\EE


\subsection{Rotation with $\al$}

The angle $\al$ is defined as
\BE 
\tan 2\al = \tan 2\be \frac{\MHpq^2 + \MZ^2}{\MHpq^2 - \MZ^2} .
\label{defalpha}
\EE
It diagonalizes (up to tadpole contributions) the matrix
$M_\rmS$ ($\sa = \Sa, \ca = \Ca$): 
\BE
\VL H \\ h \VR = D^+(\al) \VL \Pe \\ \Pz \VR
               =  \ML \ca & \sa \\ -\sa & \ca \MR \VL \Pe \\ \Pz \VR
\EE
\BEA
&& (\Pe, \Pz) \; M_\rmS \; \VL \Pe \\ \Pz \VR \non \\
&=& (\Pe, \Pz) D(\al)D^+(\al) \; M_\rmS \; D(\al)D^+(\al) 
   \VL \Pe \\ \Pz \VR \non \\
&=& (H, h) \; M^D_\rmS \; \VL H \\ h \VR
\EEA
with 
\BEA
M^D_\rmS &=& ~~\MHpq^2 \ML (\cbe \sa - \ca \sbe)^2 &
                \sbe\cbe (\sa^2 - \ca^2) + \sa\ca (\cbe^2 - \sbe^2) \\
                \sbe\cbe (\sa^2 - \ca^2) + \sa\ca (\cbe^2 - \sbe^2) &  
                       (\cbe \ca + \sa \sbe)^2 \MR \non \\
 && + \MZ^2 \ML (\ca \cbe - \sa \sbe)^2 & 
              \sbe\cbe (\sa^2 - \ca^2) - \sa\ca (\cbe^2 - \sbe^2) \\
              \sbe\cbe (\sa^2 - \ca^2) - \sa\ca (\cbe^2 - \sbe^2) &  
              (\cbe \sa + \sbe \ca)^2 \MR  \\
 && \hspace{-1em} + \bar t_1 \ML -\cbe \sbe^2 \sa^2 + 2 \sa \ca \sbe^3
                               + \cbe \ca^2 (1 + \sbe^2) &
                    \sbe^3 (\ca^2 - \sa^2) 
                               - \sa \ca \cbe (1 + 2 \sbe^2) \\
                    \sbe^3 (\ca^2 - \sa^2) 
                               - \sa \ca \cbe (1 + 2 \sbe^2) &
                    -2 \ca \sa \sbe^3
                               + \cbe (-\ca^2 \sbe^2 + \sa^2 (1 + \sbe^2))
      \MR \non \\
 && \hspace{-1em} + \bar t_2 \ML 2 \ca \sa \cbe^3 - \ca^2 \cbe^2 \sbe
                               + \sbe \sa^2 (1 + \cbe^2) &
                   \cbe^3 (\ca^2 - \sa^2)
                               + \sa \ca \sbe (1 + 2 \cbe^2) \\
                   \cbe^3 (\ca^2 - \sa^2)
                               + \sa \ca \sbe (1 + 2 \cbe^2) &
                   -2 \ca \sa \cbe^3 
                               + \sbe (\ca^2 (1 + \cbe^2) - \sa^2 \cbe^2)
      \MR \non
\EEA
Using the \refeq{defalpha} and setting the tadpoles to zero one
obtains:
\BEA
M^D_\rmS &=& ~~\MHpq^2 \ML (\cbe \sa - \ca \sbe)^2 & 0 \\
                       0 & (\cbe \ca + \sa \sbe)^2 \MR \non \\
&& + \MZ^2 \ML (\ca \cbe - \sa \sbe)^2 & 0 \\
              0 & (\cbe \sa + \sbe \ca)^2 \MR .
\EEA
The rotation with $\al$ also affects the matrix
$M^\be_\rmSP$. Defining the $(4 \times 4)$ matrix 
\BE
D^4(\al) = \ML D(\al) & 0 \\ 0 & \id \MR ,
\EE
the rotation of $M^\be_\rHiggs$ can be performed:
\BEA
&&  (\Pe, \Pz, G, A) \; M^\be_\rHiggs \; 
    \VL \Pe \\ \Pz \\ G \\ A \VR \non \\
&=&  (\Pe, \Pz, G, A) \; D^4(\al)D^{4+}(\al) \; M^\be_\rHiggs \; 
    D^4(\al)D^{4+}(\al) \; \VL \Pe \\ \Pz \\ G \\ A \VR \non \\
&=& (H, h, G, A) \; M^D_\rHiggs \; 
    \VL H \\ h \\ G \\ A \VR
\EEA
with
\BE
M^D_\rHiggs = \ML M^D_\rmS  & M^{\be\al}_\rmSP \\
                      M^{\be\al+}_\rmSP & M^D_\rmP \MR
\EE
and
\BE
M^{\be\al}_\rmSP = \bar t_A 
    \ML -\ca\sbe + \sa\cbe  & \sa \sbe + \ca\cbe \\
        \sa \sbe + \ca \cbe & \ca \sbe - \sa \cbe \MR .
\EE


\subsection{Tree-level expressions}

At tree-level all tadpoles can be set to zero. 
In the $\Pe$-$\Pz$ sector this ensures that $v_{1,2}$ are the vacuum
expectation values. In the $\Ce$-$\Cz$ sector this corresponds to a
redefinition of the phase of $m_{12}^2$ so that the phase $e^{i\xi}$
is absorbed~\cite{mhiggsCPXgen}.

One arrives at the following masses at tree-level:
\BEA
H &:& \mH^2 = \edz \KKL \MHpq^2 + \MZ^2 
                +\sqrt{(\MHpq^2 + \MZ^2)^2 
                       - 4 \MZ^2\MHpq^2 c_{2\be}^2}~\KKR \non \\
h &:& \mh^2 = \edz \KKL \MHpq^2 + \MZ^2 
                -\sqrt{(\MHpq^2 + \MZ^2)^2 
                       - 4 \MZ^2\MHpq^2 c_{2\be}^2}~\KKR \non \\
A &:& \MA^2 = \MHp^2 - \MW^2 \quad (\equiv \MHpq^2) \non \\
G &:& m_G^2 = \MZ^2 \non \\
H^\pm &:& \MHp^2~{\rm (input~value)}\non \\
G^\pm &:& m_{G^\pm}^2 = \MW^2
\EEA
The entries for the Goldstone bosons $G$ and $G^\pm$ are to be
understood in the Feynman gauge.
At tree-level there is no $\cp$ violation in the cMSSM Higgs
sector. The fields $h$ and $H$ are decoupled from the fields $A$ and
$G$.


\subsection{The scalar quark sector in the cMSSM}
\label{subsec:squarks}

The mass matrix of two squarks of the same flavor, 
$\sql$ and $\sqr$, is given by
\BE
M_{\sq} = \ML M_L^2 + \mq^2 & \mq \; \Xq^* \\
            \mq \; \Xq    & M_R^2 + \mq^2 \MR
\label{squarkmassmatrix}
\EE
with 
\BEA
M_L^2 &=& M_{\tilde Q}^2 + \MZ^2 \CZb (I_3^q - Q_q \sw^2) \non \\
M_R^2 &=& M_{\tilde Q'}^2 + \MZ^2 \CZb Q_q \sw^2 \\ \non 
\Xq &=& A_q - \mu^* \{\CTb, \tb\} ,
\label{squarksoftSUSYbreaking}
\EEA
where $\{\CTb, \tb\}$ applies for $\{ {\rm up, down} \}$-type 
squarks respectively.
In an isodoublet the $SU(2)$ symmetry enforces that $M_{\tilde Q}$ has
to be chosen equal for both squark types. The $M_{\tilde Q'}$ on the
other hand can be chosen independently for every squark type.
In the scalar quark sector of the cMSSM $N_q + 1$ phases are
present, one for each $A_q$ and one 
for $\mu$, i.e.\ $N_q + 1$ new parameters appear. As an abbreviation it
will be used
\BE
\phi_q = {\rm arg}\KL \Xq \KR .
\EE
As an independent parameter one can trade 
${\rm arg}\KL A_q \KR \equiv \phi_{A_q}$ for $\phi_q$. \\
The squark mass eigenstates are obtained by the rotation
\BE
\VL \sqe \\ \sqz \VR = S^{\sq} \VL \sql \\ \sqr \VR
\EE
with
\BE
S^{\sq} = \ML \ctq & \stq^* \\ -\stq & \ctq^* \MR 
         = \ML  e^{ i\phi_q/2} |\ctq| & e^{-i\phi_q/2} |\stq| \\
               -e^{ i\phi_q/2} |\stq| & e^{-i\phi_q/2} |\ctq| \MR
               ,\quad
S^{\sq+} S^{\sq} = \id~,
\EE
where the matrix with $\phi_q \to 0$ diagonalizes 
$M_{\sq_{\Big|\Xq \to |\Xq|}}$.
The mass eigenvalues are given by
\BE
m_{\tilde q_{1,2}}^2 = \mq^2 
  + \edz \KKL M_L^2 + M_R^2 
           \mp \sqrt{( M_L^2 - M_R^2)^2 + 4 \mq^2 |\Xq|^2}~\KKR ,
\EE
independent of the phase of $\Xq$. 
The unrotated squark mass matrix can now be expressed in terms of the
physical parameters $\msqe, \msqz$ and the $\sq$ mixing angle:
\BE
M_{\sq} = \ML \ctq\ctq^* \msqe^2 + \stq\stq^* \msqz^2 &
              \stq^* \ctq^* (\msqe^2 - \msqz^2) \\
              \stq   \ctq   (\msqe^2 - \msqz^2) &
              \stq\stq^* \msqe^2 + \ctq\ctq^* \msqz^2 \MR .
\label{squarkmassmatrixphys}
\EE


\section{Calculation of the renormalized self-energies}

\subsection{Renormalization}
\label{subsec:ren}

The renormalization is performed as follows:
\BEA
\MHp^2 &\to& \MHp^2 + \de\MHp^2 \non \\
\MW^2 &\to& \MW^2 + \de\MW^2 \non \\
\MZ^2 &\to& \MZ^2 + \de\MZ^2 \non \\
t_x &\to& t_x + \de t_x, \quad x = 1, 2, A \non \\
\tb &\to& \tb + \de\tb \non \\
\cHe &\to& Z_{H_1}^{1/2} \cHe \non \\
\cHz &\to& Z_{H_2}^{1/2} \cHz
\EEA
The counterterm for the $A$ tadpole can be understood as the effect of
re normalizing the phase $\xi$ of $\cHz$.

\bigskip
In the following we will concentrate on the contributions that are
relevant for the leading $\mt^4$ corrections (or any corrections of
the type $\sim \mf^4$) for the masses of the
neutral Higgs bosons. There, only 
$\de\MHp^2$ and $\de t_x, x = 1,2,A$, enter (see also
\citere{mhiggslong}). 

\smallskip
\noindent
The renormalized $H^\pm$ self-energy is then given by
\BE
\re\hSi_{H^\pm}(0) = \Si_{H^\pm}(0) - \de\MHp^2 ,
\EE
the renormalized tadpoles are given by
\BE
\hat t_x = T_x + \de t_x, \quad x = 1, 2, A ,
\EE
$T_x$ represents the \onel\ contribution to $t_x$.

\smallskip
\noindent
The on-shell renormalization conditions are imposed:
\BEA
\re\hSi_{H^\pm}(0) &=& 0 \; , \\
\hat t_x &=& 0 \; .
\EEA
This results in the on-shell renormalization constants
\BEA
\de\MHp^2 &=& \Si_{H^\pm}(0) \; , \\
\de t_x  &=& -T_x \; .
\EEA 
Since the charged Higgs boson is renormalized on-shell, its mass does
not receive higher-order corrections.


\subsection{Renormalized self-energies}

With the on-shell renormalization constants derived in
\refse{subsec:ren} the renormalized neutral Higgs boson self-energies
read: 
\BEA
\label{hhren}
\hSih(0) &=& \Sih(0) - \de\MHp^2 (\ca\cbe + \sa\sbe)^2 \non\\
 && + T_1 \frac{e}{2\sw\MW} 
      (-2 \ca \sa \sbe^3 + \cbe (-\ca^2\sbe^2 + \sa^2 (1 + \sbe^2))) \non\\
 && + T_2 \frac{e}{2\sw\MW} 
      (-2 \ca \sa \cbe^3 + \sbe (\ca^2 (1 + \cbe^2) - \sa^2\cbe^2)) \\
\label{HHren}
\hSiH(0) &=& \SiH(0) - \de\MHp^2 (\sa\cbe - \ca\sbe)^2 \non \\
 && + T_1 \frac{e}{2\sw\MW}
      (-\cbe \sa^2 \sbe^2 + 2 \sa\ca\sbe^3 + \ca^2 \cbe (1 + \sbe^2)) \non\\
 && + T_2 \frac{e}{2\sw\MW}
      (2 \sa\ca\cbe^3 - \ca^2\cbe^2\sbe + (1 + \cbe^2) \sa^2\sbe) \\
\label{hHren}
\hSihH(0) &=& \SihH(0) - \de\MHp^2 (\sbe\cbe (\sa^2 - \ca^2)
                                   + \sa\ca (\cbe^2 - \sbe^2)) \non \\
 && + T_1 \frac{e}{2\sw\MW} 
      (\sbe^3 (\ca^2 - \sa^2) - \sa\ca\cbe (1 + 2 \sbe^2)) \non \\
 && + T_2 \frac{e}{2\sw\MW}
      (\cbe^3 (\ca^2 - \sa^2) + \sa\ca\sbe (1 + 2 \cbe^2)) \\
\label{AAren}
\hSi_{AA}(0) &=& \Si_{AA}(0) - \de\MHp^2 \\
\label{GGren}
\hSi_{GG}(0) &=& \Si_{GG}(0) + \frac{e}{2\sw\MW} (-T_1 \cbe - T_2 \sbe) \\
\label{AGren}
\hSi_{AG}(0) &=& \Si_{AG}(0) + \frac{e}{2\sw\MW} (-T_1 \sbe + T_2 \cbe) \\
\label{hAren}
\hSi_{hA}(0) &=& \Si_{hA}(0) + T_A \frac{e}{2\sw\MW} (-\ca\sbe + \sa\cbe) \\
\label{HAren}
\hSi_{HA}(0) &=& \Si_{HA}(0) + T_A \frac{e}{2\sw\MW} (-\sa\sbe - \ca\cbe) \\
\label{hGren}
\hSi_{hG}(0) &=& \Si_{hG}(0) + T_A \frac{e}{2\sw\MW} (-\sa\sbe - \ca\cbe) \\\
\label{HGren}
\hSi_{HG}(0) &=& \Si_{HG}(0) + T_A \frac{e}{2\sw\MW} (-\sa\cbe + \ca\sbe) 
\EEA


\subsection{Evaluation of $\mt^4$ contributions}
\label{subsec:mt4eval}

\begin{figure}[ht!]
\vspace{2.5em}
\begin{center}
\mbox{
\psfig{figure=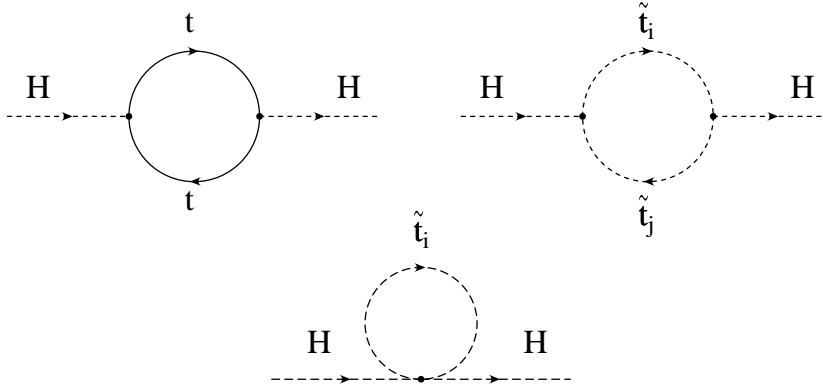,width=12cm,bbllx=150pt,bblly=630pt,
                                      bburx=450pt,bbury=730pt}}
\end{center}
\caption[]{\it\footnotesize 
Generic Feynman diagrams for the $\mt^4$ contributions to Higgs
self-energies. 
}
\label{fig:fdol}
\end{figure}

\begin{figure}[ht!]
\vspace{2.5em}
\begin{center}
\mbox{
\psfig{figure=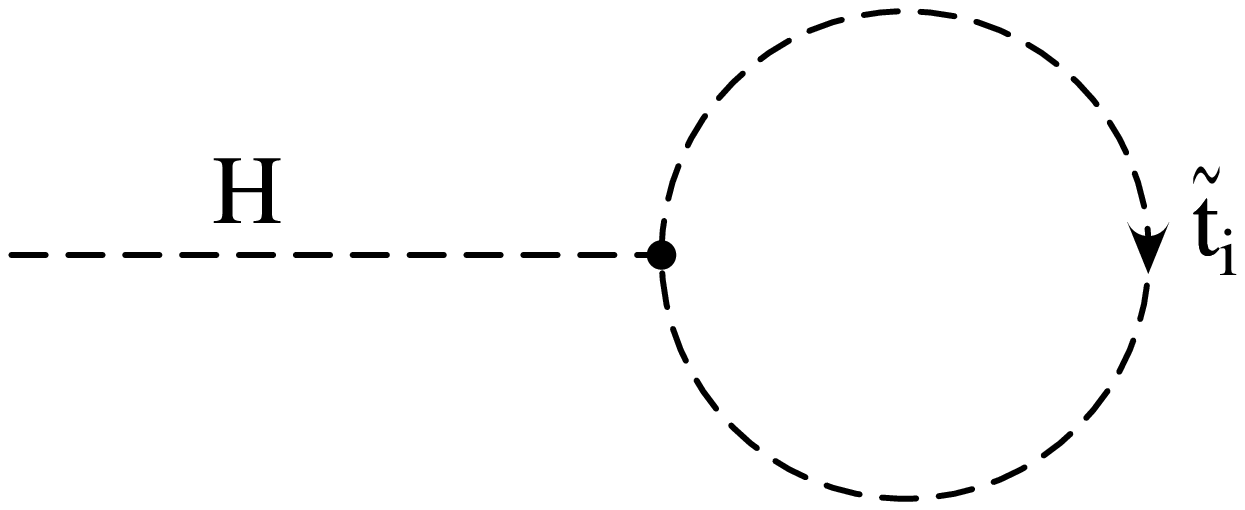,width=3.5cm,bbllx=150pt,bblly=370pt,
                                      bburx=450pt,bbury=470pt}
\hspace{2cm}
\psfig{figure=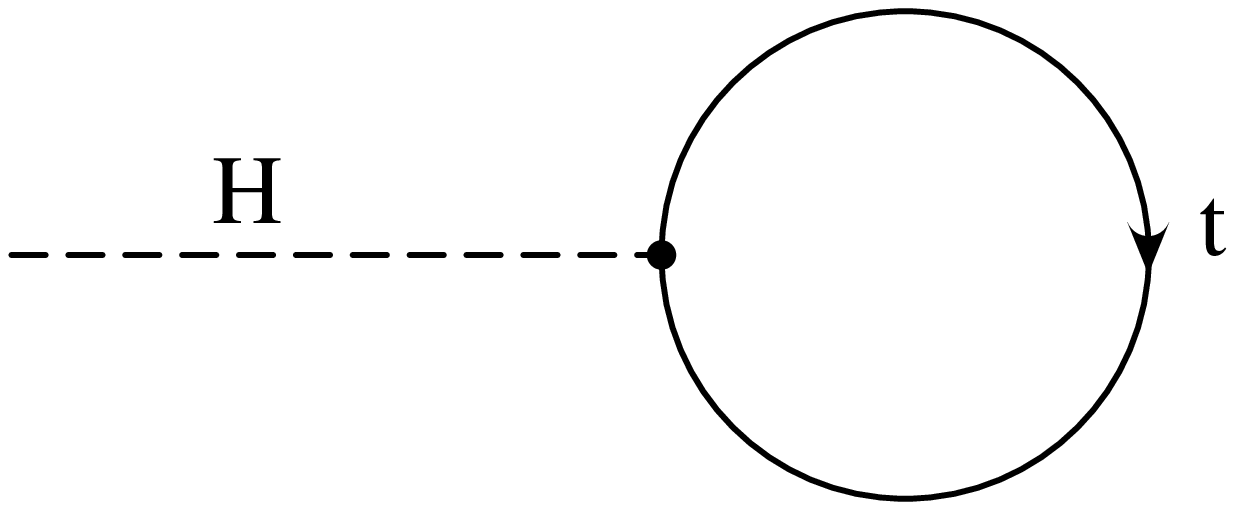,width=3.5cm,bbllx=150pt,bblly=370pt,
                                      bburx=450pt,bbury=470pt}
}
\end{center}
\caption[]{\it\footnotesize 
Generic Feynman diagrams for the $\mt^4$ contributions to Higgs tadpoles.
}
\label{fig:fdol_tp}
\end{figure}

For the evaluation of the leading $\mt^4$ corrections in the
Feynman-diagrammatic (FD) approach the diagrams
shown in \reffi{fig:fdol} have to be evaluated for all self-energies
$\Si_{st}(0), st = hh, HH, hH, AA, GG, AG, hA, HA, hG, HG$. Concerning the
tadpole contributions the diagrams of \reffi{fig:fdol_tp} have to be
considered. 

For sake of simplicity we now switch back to the $\Pe$-$\Pz$ basis
(i.e. $\al = 0$), where the results have a much simpler form. The
corresponding results in the $h$-$H$ basis can be obtained by the
rotation
\BEA
\hSih &=& \SQa \hSi_{\Pe} + \CQa \hSi_{\Pz} - 
              2 \Sa \Ca \hSi_{\PePz} \non \\
\hSiH &=& \CQa \hSi_{\Pe} + \SQa \hSi_{\Pz} + 
              2 \Sa \Ca \hSi_{\PePz} \non \\
\hSihH &=& - \Sa \Ca \KL \hSi_{\Pe} - \hSi_{\Pz} \KR + 
              (\CQa - \SQa) \hSi_{\PePz} \non \\
\hSi_{hA} &=& -\Sa \hSi_{\Pe A} + \Ca \hSi_{\Pz A} \non \\
\hSi_{HA} &=& \Ca \hSi_{\Pe A} + \Sa \hSi_{\Pz A} \non \\
\hSi_{hG} &=& -\Sa \hSi_{\Pe G} + \Ca \hSi_{\Pz G} \non \\
\hSi_{HG} &=& \Ca \hSi_{\Pe G} + \Sa \hSi_{\Pz G} .
\label{higgsserotation}
\EEA

The $\mt^4$ corrections have been obtained using the program 
\fa~3~\cite{feynarts}, employing the recently completed MSSM model 
file~\cite{mssmmodelfile}%
\footnote{
Only the non-SM counterterms had to be added.
}%
. Details about the calculations with \fa\ can be found in
\citere{feynartsuse}.
In the approximation of the leading $\mt^4$ corrections (and with 
$\mb = 0$) the result for
the renormalized self-energies of \refeq{hhren} - (\ref{HGren}) reads:
\BEA
\label{p1p1se}
\hSi_{\Pe\Pe}(0) &=& \frac{3 \, e^2 \MZ^2}{32 (\MW^2 - \MZ^2) \pi^2 \sbe^2}
                  \frac{\mt^2}{\MW^2}\;\times \non \\
&&  \Bigg\{
  (\stt^2 \ctt^2 \mu^2 + \stt^{*2} \ctt^{*2} \mu^{*2})\, g(\mste, \mstz)
  -\Deca \sbe^2 \Bigg\}\\
\hSi_{\Pz\Pz}(0) &=& \frac{3 \, e^2 \MZ^2}{32 (\MW^2 - \MZ^2) \pi^2 \sbe^2}
                  \frac{\mt^2}{\MW^2} \Bigg\{ -\Deca \cbe^2 \;
   -2 \mt^2 \log \KL \frac{\mt^4}{\mste^2 \mstz^2} \KR \non \\
&& + \Bigg[
  \frac{\cbe^2}{\sbe^2} (\stt^2 \ctt^2 \mu^2 + \stt^{*2} \ctt^{*2} \mu^{*2})
 + 2 \frac{\cbe}{\sbe} \frac{\mste^2 - \mstz^2}{\mt}
     ( \stt \ctt \mu + \stt^* \ctt^* \mu^*) \stt\stt^* \ctt\ctt^* \non \\
&&   + 2 \stt^2 \stt^{*2} \ctt^2 \ctt^{*2} \frac{(\mste^2 - \mstz^2)^2}{\mt^2}
     \Bigg] \, g(\mste, \mstz) \non \\
&& + \Bigg[ 2 \mt \frac{\cbe}{\sbe} 
            ( \stt \ctt \mu + \stt^* \ctt^* \mu^*)
           + 4 \stt\stt^* \ctt\ctt^* (\mste^2 - \mstz^2) \Bigg]
     \log \KL \frac{\mste^2}{\mstz^2} \KR \Bigg\} \\
\hSi_{\PePz}(0) &=& - \frac{3 \, e^2 \MZ^2}{32 (\MW^2 - \MZ^2) \pi^2 \sbe^2}
                  \frac{\mt^2}{\MW^2} \Bigg\{ -\Deca \sbe\cbe \; +\Bigg[
   \frac{\cbe}{\sbe} (\stt^2 \ctt^2 \mu^2 + \stt^{*2} \ctt^{*2} \mu^{*2})
   \non \\
&& + ( \stt \ctt \mu + \stt^* \ctt^* \mu^*)
      \stt\stt^* \ctt\ctt^* \frac{\mste^2 - \mstz^2}{\mt} \Bigg]
     \, g(\mste, \mstz) \non \\
&&  + ( \stt \ctt \mu + \stt^* \ctt^* \mu^*) \mt
     \log \KL \frac{\mste^2}{\mstz^2} \KR \Bigg\} \\
\label{aase}
\hSi_{AA}(0) &=& - \frac{3 \, e^2 \MZ^2}{32 (\MW^2 - \MZ^2) \pi^2 \sbe^2}
                  \frac{\mt^2}{\MW^2} \; \Deca 
                  \quad \KL \equiv \Si_{AA}(0) - \Si_{H^\pm}(0) \KR \\
\hSi_{H^\pm}(0) &=& 0 \quad({\rm by~renormalization}) \\
\hSi_{GG}(0) &=& 0 \\
\hSi_{AG}(0) &=& 0 \\
\hSi_{\Pe A}(0) &=& \frac{3i \, e^2 \MZ^2}{64 (\MW^2 - \MZ^2) \pi^2 \sbe^3}
                  \frac{\mt^2}{\MW^2}
  (\stt^2 \ctt^2 \mu^2 - \stt^{*2} \ctt^{*2} \mu^{*2}) \, g(\mste, \mstz) \\
\hSi_{\Pz A}(0) &=& -\frac{3i \, e^2 \MZ^2}{64 (\MW^2 - \MZ^2) \pi^2 \sbe^3}
                  \frac{\mt^2}{\MW^2} \Bigg\{ \Bigg[
   \frac{\cbe}{\sbe} (\stt^2 \ctt^2 \mu^2 - \stt^{*2} \ctt^{*2} \mu^{*2})
   \non \\
&&  + 2 \frac{\mste^2 - \mstz^2}{\mt}
     (\stt \ctt \mu - \stt^{*} \ctt^{*} \mu^{*}) \stt\stt^*\ctt\ctt^* \Bigg]
     \, g(\mste, \mstz) \non \\
&& + 2 \mt (\stt \ctt \mu - \stt^{*} \ctt^{*} \mu^{*}) 
     \log \KL \frac{\mste^2}{\mstz^2} \KR \Bigg\} \\
\hSi_{\Pe G}(0) &=& 0 \\
\hSi_{\Pz G}(0) &=& 0 
\EEA
with 
\BEA
g(x, y) &=& 2 - \frac{x^2 + y^2}{x^2 - y^2} \log\KL \frac{x^2}{y^2} \KR \non\\
\label{DeltaCA}
\Deca &\equiv& \frac{1}{\sbe^2} 
               \KKL \stt^2 \ctt^2 \mu^2 
                    - 2 \stt \stt^* \ctt \ctt^* \mu \mu^*
                    + \stt^{*2} \ctt^{*2} \mu^{*2} \KKR \non \\
&& - \frac{\msbl^2 \mt^2 \mu \mu^*}
          {2 \sbe^2 (\msbl^2 - \mste^2) (\msbl^2 - \mstz^2)}
   \log\KL \frac{\msbl^4}{\mste^2\mstz^2} \KR \non \\
&& + \frac{1}{2 \sbe^2} \log\KL \frac{\mste^2}{\mstz^2} \KR 
  \Bigg\{ -\frac{\mste^2 + \mstz^2}{\mste^2 - \mstz^2}
          \KL \stt^2 \ctt^2 \mu^2 + \stt^{*2} \ctt^{*2} \mu^{*2} \KR \non \\
&&       + \mu \mu^* \Bigg[ 2 (\stt \stt^* - \ctt \ctt^*)
                          -\frac{2}{\mste^2 - \mstz^2}
                           \KL \stt^2 \stt^{*2} \mste^2 
                              +\ctt^2 \ctt^{*2} \mstz^2 \KR \non \\
&&             + \msbl^2 \KL \frac{\ctt \ctt^*}{\msbl^2 - \mstz^2}
                            -\frac{\stt \stt^*}{\msbl^2 - \mste^2} \KR\Bigg]
  \Bigg\} \\
\msbl^2 &\equiv& \ctt \ctt^* \mste^2 + \stt \stt^* \mstz^2 - \mt^2
\label{msbot}
\EEA

As expected, $\hSi_{hG}(0) = \hSi_{HG}(0) = \hSi_{AG}(0) = \hSi_{GG}(0) = 0$, 
i.e.\ the Goldstone boson $G$ decouples~\cite{mhiggsCPXgen}. 
In order to show the finiteness
of $\hSi_{st}, st = \Pe\Pe, \Pz\Pz, \Pe\Pz, AA, \Pe A, \Pz A$ 
it was necessary to
employ the $SU(2)$ symmetry in the scalar quark sector, see
\refse{subsec:squarks}. In the 
simplified case of the leading $\mt^4$ corrections and $\mb = 0$
(i.e.\ no mixing in the $\Sbot$~sector) only the left-handed scalar
bottom quark, $\SbotL$ contributes, where its mass is given by
\refeq{msbot}. 

Exactly analogous expressions have been obtained for the leading
$\mb^4$ corrections (with $\mst \leftrightarrow \msb$ and 
$\sbe \leftrightarrow \cbe$ (except in the $\Deca$ prefactor)), 
which can be relevant for large $\tb$. Analogous to the $\mb^4$
corrections also the corresponding $\mtau^4$ contributions 
(up to the color factor and with $\msb \to \mstau$) have been evaluated.

\bigskip
A main difference compared to the RG improved EP
approach as presented in \citere{mhiggsCPXRG1} is
the validity of the result as a function of the $\Stop$ sector
parameters. Since the FD result is obtained directly in terms of the
physical parameters in the squark sector, the results of the FD
approach are valid for arbitrary 
mixing in the $\Stop$ sector, whereas the RG method is restricted to
$(\mstz^2 - \mste^2)/(\mstz^2 + \mste^2) \lsim 1/2$.


\subsection{Corrections beyond \onel\ order}
\label{subsec:mhiggs2lcorr}

Since it is known in the case of vanishing complex phases that the \twol\
corrections to the neutral Higgs boson masses can be large, for the
further numerical examples and comparisons as presented in
\refse{sec:numeval}, the leading contributions at
\order{\gf\als\mt^4} and \order{\gf^2\mt^6} are taken into account. For
sake of simplicity, up to now the \twol\ corrections are taken over 
from the $\cp$-conserving case. The leading corrections then only
affect $\hSi_{\Pz\Pz}(0)$ and are valid for arbitrary Higgs sector
parameters. They are given 
by~\cite{mhiggsRG,mhiggslle,mhiggsRG2}
\BEA 
\label{mh2twolooptop}
\hSi_{\Pz\Pz}^{2,\al\als}(0) &=&
    \frac{\gf\wz}{\pi^2} \frac{\als}{\pi}\; \frac{\mtms^4}{\SQb}
      \Biggl[ 4 + 3 \log^2\lmtmsms + 2 \log\lmtmsms 
             -6 \frac{\Xt}{\ms} \non \\
 && {}  
     - \frac{\Xt^2}{\ms^2} \KKKL 3 \log\lmtmsms +8 \KKKR 
      +\frac{17}{12} \frac{\Xt^4}{\ms^4} \Biggr] \\ 
\label{mh2yuk}
\hSi_{\Pz\Pz}^{2,\al^2}(0) &=& -\frac{9}{16\pi^4} G_F^2 \frac{\mtms^6}{\SQb}
               \KKL \tilde{X} t + t^2 \KKR~, \\
\tilde{X} &=& \Bigg[
                \KL \frac{\mstz^2 - \mste^2}{4 \mtms^2} \sinQZtt \KR^2
                \KL 2 - \frac{\mstz^2 + \mste^2}{\mstz^2 - \mste^2}
                      \log\KL \frac{\mstz^2}{\mste^2} \KR \KR \non\\
            && {} 
               + \frac{\mstz^2 - \mste^2}{2 \mtms^2} \sinQZtt
                      \log\KL \frac{\mstz^2}{\mste^2} \KR \Bigg], 
              \quad t = 
              \frac{1}{2} \log \KL \frac{\mste^2 \mstz^2}{\mtms^4} \KR
          \non .
\EEA
$\ms$ has to be chosen according to
\BE
\label{msfkt}
\ms = \KKKL \begin{array}{l@{\quad:\quad}l}
            \sqrt{\msq^2 + \mt^2} & \MstL = \MstR = \msq \\
            \KKL \MstL^2 \MstR^2 + \mt^2 (\MstL^2 + \MstR^2) +
                     \mt^4 \KKR^\frac{1}{4} &
                                    \MstL \neq \MstR
            \end{array} \right. 
\EE

\noindent
and $\mtms$ denotes the running top quark mass, $\mtms = \mtms(\mt)$.
$\MstL, \MstR$ correspond to $M_{\tilde Q}, M_{\tilde Q'}$ in
\refeq{squarksoftSUSYbreaking} respectively. Contrary to the presented
\onel\ result for $\hSi_{\Pz\Pz}$, \refeq{mh2twolooptop} is valid only for
not too large mass splitting between the two $\Stop$~mass
eigenstates, but still gives a rather good approximation for a large
part of the MSSM parameter space~\cite{mhiggslle}.  The full result in
\citere{mhiggsCPXFD}, however, will be 
obtained in terms of the physical parameters and thus be valid for
arbitrary mixing in the $\Stop$~sector.
Also \refeq{mh2yuk} is valid for not too large mass splitting in the
$\Stop$~sector~\cite{mhiggsRG,mhiggsRG2}. However, since the numerical
effect of the correction in \refeq{mh2yuk} is at the $\sim 2 \gev$
level~\cite{mhiggslong}, this additional uncertainty is
neglected. Furthermore, \refeq{mh2yuk} has been obtained in the
\msbar\ scheme, while all other corrections in this paper are
evaluated in the on-shell scheme. The corresponding uncertainty is only
of \order{\al^2\als} and expected to be below $\sim 1 \gev$ and
therefore neglected.


\section{The neutral MSSM Higgs sector}

\subsection{The Higgs boson masses}

In this section, for sake of completeness, we review the derivation of
the Higgs boson masses from the calculated higher-order Higgs boson
self-energies. Since in the approximation used in
\refse{subsec:mt4eval} the external momentum has been set to zero,
this step of the evaluation is equal to the EP
approach~\cite{mhiggsCPXEP1,mhiggsCPXEP2,mhiggsCPXRG1}. In the full FD
calculation~\cite{mhiggsCPXFD} the momentum dependence, however, is
included, which can lead to corrections of $1-2 \gev$.

\smallskip
Since the Goldstone boson $G$ decouples, see \refse{subsec:mt4eval},
the fields $\Pe$, $\Pz$ and $A$ 
form a closed subspace that can be evaluated on its own. The masses at
higher order can be obtained from the diagonalization of the matrix
\BEA
M_\cp &=& \MLd M_{11} & M_{12} & M_{13} \\
               M_{21} & M_{22} & M_{23} \\
               M_{31} & M_{32} & M_{33} \MR \non \\
      &=& \MLd \MHpq^2 -\hSi_{AA}(0)&          -\hSi_{\Pe A}(0)   & 
                                               -\hSi_{\Pz A}(0)   \\
                   -\hSi_{\Pe A}(0) & \mpe^2   - \hSi_{\Pe\Pe}(0) &
                                      \mpez^2  - \hSi_{\PePz}(0)  \\
                   -\hSi_{\Pz A}(0) & \mpez^2  - \hSi_{\PePz}(0)  &
                                      \mpz^2   - \hSi_{\Pz\Pz}(0)
          \MR .
\EEA
The diagonalization is performed with the help of the $(3 \times 3)$
orthogonal matrix $D^3$:
\BEA
&& \KL A, \Pe, \Pz \KR \; M_\cp \; \VL A \\ \Pe \\ \Pz \VR \non \\
&=& \KL A, \Pe, \Pz \KR \; D^3 D^{3+} \; M_\cp \; 
     D^3 D^{3+} \; \VL A \\ \Pe \\ \Pz \VR \non \\
&=& \KL \Hd, \Hz, \He \KR \; M_\cp^D \; \VL \Hd \\ \Hz \\ \He \MR
\EEA
with 
\BE
M_\cp^D = \MLd \mHd^2 & 0 & 0 \\
               0 & \mHz^2 & 0 \\
               0 & 0 & \mHe^2 \MR \; , \quad
\mHd \ge \mHz \ge \mHe .
\label{massordering}
\EE

\bigskip
The numerical evaluation of $M_\cp^D$ and $D^3$ has been presented
e.g. in \citere{mhiggsCPXRG1} and is also listed here for completeness.
The eigenvalues of $M_\cp$ are given by
\BEA
e_1 &=& -\ed{3} r + 2 \sqrt{-p/3}\; \cos\KL \frac{\varphi}{3} \KR , \non \\ 
e_2 &=& -\ed{3} r + 2 \sqrt{-p/3}\; \cos\KL \frac{\varphi}{3} 
                                         + \frac{2\pi}{3} \KR , \non \\
e_3 &=& -\ed{3} r + 2 \sqrt{-p/3}\; \cos\KL \frac{\varphi}{3} 
                                         - \frac{2\pi}{3} \KR ,
\EEA
with 
\BEA
p = \frac{3s - r^2}{3} , \quad
q = \frac{2 r^3}{27} - \frac{r s}{3} + t , \quad
\varphi = \arccos \KL -\frac{q}{2 \sqrt{-p^3/27}} \KR
\EEA
and
\BEA
r = -\mbox{Tr} \KL M_\cp \KR , \quad
s = \edz \KKL \mbox{Tr}^2 \KL M_\cp \KR 
              - \mbox{Tr} \KL M_\cp^2 \KR \KKR , \quad
t = - \mbox{Det} \KL M_\cp \KR . 
\EEA

\noindent
The rotation matrix $D^3$ can be obtained as
\BE
D^3 = \MLd |x_1|/\De_1 &  x_2 /\De_2 &  x_3 /\De_3 \\
            y_1 /\De_1 & |y_2|/\De_2 &  y_3 /\De_3 \\
            z_1 /\De_1 &  z_2 /\De_2 & |z_3|/\De_3 \MR , \quad
\De_i = \sqrt{x_i^2 + y_i^2 + z_i^2} , 
\EE
where
\BEA
x_1 &=& \mbox{Det} \ML M_{22} - \mHd^2 & M_{23} \\ 
                       M_{32} & M_{33} - \mHd^2 \MR \non \\
y_2 &=& \mbox{Det} \ML M_{11} - \mHz^2 & M_{13} \\ 
                       M_{31} & M_{33} - \mHz^2 \MR \non \\
z_3 &=& \mbox{Det} \ML M_{11} - \mHe^2 & M_{12} \\ 
                       M_{21} & M_{22} - \mHe^2 \MR \non \\
x_2 &=& \mbox{Det} \ML M_{13} & M_{12} \\ 
                       M_{33} - \mHz^2 & M_{32} \MR 
         \times \mbox{sign}(y_2) \non \\
x_3 &=& \mbox{Det} \ML M_{12} & M_{13} \\ 
                       M_{22} - \mHe^2 & M_{23} \MR 
         \times \mbox{sign}(z_3) \non \\
y_1 &=& \mbox{Det} \ML M_{23} & M_{21} \\ 
                       M_{33} - \mHd^2 & M_{31} \MR 
         \times \mbox{sign}(x_1) \non \\
y_3 &=& \mbox{Det} \ML M_{13} & M_{11} - \mHe^2 \\ 
                       M_{23} & M_{21} \MR 
         \times \mbox{sign}(z_3) \non \\
z_1 &=& \mbox{Det} \ML M_{21} & M_{22} - \mHd^2 \\ 
                       M_{31} & M_{32} \MR 
         \times \mbox{sign}(x_1) \non \\
z_2 &=& \mbox{Det} \ML M_{12} & M_{11} - \mHz^2 \\ 
                       M_{32} & M_{31} \MR 
         \times \mbox{sign}(y_2)
\EEA


\subsection{The Higgs boson couplings}

Again we follow the prescriptions as given in \citere{mhiggsCPXRG1}.
Taking complex phases into account, all three neutral Higgs bosons
are composed of a $\cp$-even part, thus all three Higgs bosons can
couple to two gauge boson, $VV = ZZ, W^+W^-$. The coupling normalized
to the SM value is given by
\BE
g_{H_iVV} = \cbe D^3_{2,4-i} + \sbe D^3_{3,4-i} \quad .
\EE
The coupling of two Higgs bosons to a $Z$~boson, normalized to
the SM value, is given by
\BE
g_{H_iH_jZ} = D^3_{1,4-i} \KL \cbe D^3_{3,4-j} - \sbe D^3_{2,4-j} \KR
            - D^3_{1,4-j} \KL \cbe D^3_{3,4-i} - \sbe D^3_{2,4-i} \KR .
\EE
The Bose symmetry that forbids any anti-symmetric derivative coupling
of a vector particle to two identical real scalar fields is respected, 
$g_{H_iH_iV} = 0$.

Finally the decay width of the $H_i$ to SM fermions can be obtained
from the decay width of the SM Higgs boson by multiplying it with 
\BE
\KKL \KL g_{H_iff}^\rmS \KR^2 + \KL g_{H_iff}^\rmP \KR^2 \KKR,
\EE
with
\BEA
&& g_{H_iuu}^\rmS = D^3_{3,4-i}/\sbe, \quad 
   g_{H_iuu}^\rmP = D^3_{1,4-i}\; \cbe/\sbe \\
&& g_{H_idd}^\rmS = D^3_{2,4-i}/\cbe, \quad 
   g_{H_idd}^\rmP = D^3_{1,4-i}\; \sbe/\cbe 
\EEA
for up- and down-type quarks respectively.


\subsection{The special case of vanishing phases}

In the $\cp$ conserving case, e.g.\ for the leading $\mt^4$
corrections $\phi_t = \phi_\mu = 0$, the $\cp$-even Higgs bosons
(denoted as $h$ and $H$ with $\mh \le \mH$) and $\cp$-odd
Higgs boson (denoted as~$A$) do not mix. The unrotated mass matrix is
then given by
\BE
M_\cp = \MLd \MHpq^2 -\hSi_{AA}(0)&                        0    & 
                                                           0    \\
                               0  & \mpe^2   - \hSi_{\Pe\Pe}(0) &
                                    \mpez^2  - \hSi_{\PePz}(0)  \\
                               0  & \mpez^2  - \hSi_{\PePz}(0)  &
                                    \mpz^2   - \hSi_{\Pz\Pz}(0)
     \MR 
\EE
where the square of the $\cp$-odd Higgs boson mass is given by 
$\MA^2 = \MHpq^2 -\hSi_{AA}(0)$. For a large part of the MSSM
parameter space the mass ordering for the three Higgs boson masses is
given as $\mH \ge \MA \ge \mh$, i.e.
\BE
M_\cp^D = \MLd \mH^2 & 0 & 0 \\
          0 & \MA^2 & 0 \\
          0 & 0 & \mh^2 \MR \quad \mbox{and} \quad
D^3 = \MLd 0 & 1 & 0 \\
           \sa & 0 & \ca \\
           \ca & 0 & -\sa \MR .
\EE
The mass ordering in \refeq{massordering} can thus imply that in the limit
of vanishing phases $H_2$ is the $\cp$-odd Higgs boson.


\section{Numerical examples and comparison with other approaches}
\label{sec:numeval}

The results obtained in \refse{subsec:mt4eval}, \refeqs{p1p1se} -
(\ref{DeltaCA}), have been compared analytically with the
corresponding results presented in \citere{mhiggsCPXEP2} 
(eqs.~(11) - (18c)). \citere{mhiggsCPXEP2} calculates the leading
corrections to the Higgs boson mass matrix in the EP approach. In the
approximation of zero external momentum as applied in
\refse{subsec:mt4eval}, the leading $\mt^4$ corrections as presented
in \refeqs{p1p1se} - (\ref{DeltaCA}) should therefore agree with the
corresponding results in \citere{mhiggsCPXEP2}. Differences due to
different renormalization schemes are only expected from \twol\ order
on, see \citere{bse,mhiggsRGFD3}. Complete analytical agreement
between the two results is found, if the correction to $\MHaa$ ,the
$A$~boson propagator in eq.~(11) of \citere{mhiggsCPXEP2} is identified
with our renormalized $A$~boson self-energy, $\hSi_{AA}$, given in
\refeq{aase}. $\hSi_{AA}$ exhibits an additional term compared to the
correction to $\MHaa$,
arising from the fact that in \citere{mhiggsCPXEP2} the charged Higgs
boson sector has been neglected, while in our approach $\MHp$ is chosen as
an input parameter, thus introducing $\Si_{H^\pm}$ into the result.
Therefore, this difference only reflects the fact of a different
choice of input parameters. A similar observation has already been
made in \citere{mhiggsCPXEP2}, while comparing with
\citere{mhiggsCPXRG1} (where also analytical agreement in the
appropriate limits has been found.) 

\smallskip
In the following subsections some numerical examples are presented and
compared to 
results obtained in the RG improved EP calculation. The examples are
based on the results given in
\refses{subsec:mt4eval},~\ref{subsec:mhiggs2lcorr}. They are meant
to illustrate the possible effects of complex phases in the MSSM. For
a phenomenological analysis, however constraints on $\cp$-violating
parameters from experimental bounds on electric dipole moments (EDMs) 
have to be taken into account, see \refse{sec:feynhiggsfastc}. 
On the other hand, the bounds from
EDMs can easily be evaded by making the first two generations
sufficiently heavy~\cite{heavy12}.
A more detailed phenomenological analysis of the FD results, including
the full \onel\ and leading \twol\ corrections in the cMSSM to the
Higgs boson self-energies, and taking into account 
all existing experimental constraints can be found in
\citere{mhiggsCPXFD}.

\subsection{Higgs boson masses}

\begin{figure}[htb!]
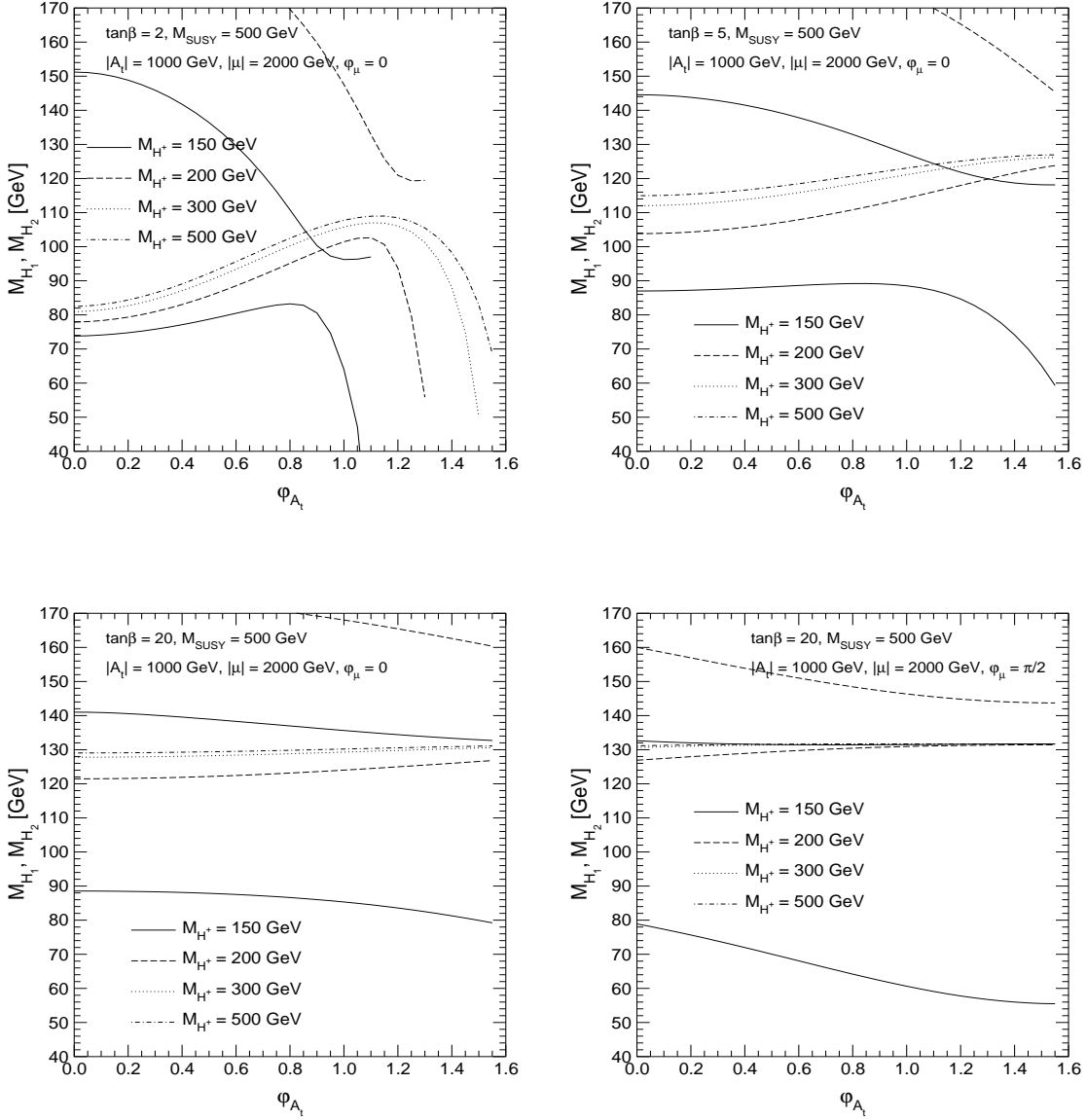

\vspace{5em}
\begin{center}
\mbox{
\epsfig{figure=mhcMSSM21.bw.eps,width=7cm,height=7.0cm} 
\hspace{1em}
\epsfig{figure=mhcMSSM22.bw.eps,width=7cm,height=7.0cm}}

\vspace{3em}
\mbox{
\epsfig{figure=mhcMSSM23.bw.eps,width=7cm,height=7.0cm} 
\hspace{1em}
\epsfig{figure=mhcMSSM26.bw.eps,width=7cm,height=7.0cm}}
\end{center}
\vspace{1em}
\caption[]{
The two lightest neutral Higgs boson masses are shown as a function of
$\phiat$ for different values of $\MHp$. 
In the first three plots $\phimu$ is set to zero and $\tb$ is chosen
as $2, 5, 20$. In the last plot $\phimu = \pi/2$ and $\tb = 20$ is
taken. The other parameters are $\msusy = 500 \gev, |\At| = 1000 \gev$
and $|\mu| = 2000 \gev$.
}
\label{fig:mh1mh2_phiat}
\vspace{1em}
\end{figure}

In \reffi{fig:mh1mh2_phiat} the two lightest Higgs boson masses,
$\mHe$ and $\mHz$, are shown as a function of the phase of the
trilinear coupling in the $\Stop$ sector, $\phiat$. The soft
SUSY-breaking parameters are chosen to emphasize the effect of the
$\cp$-violating phases, $\msusy = 500 \gev$, $|\At| = 1000 \gev$ and
$|\mu| = 2000 \gev$. The phase of $\mu$ is chosen to be zero, except
for the lower right plot, where it is set to $\phimu = \pi/2$. The
phases in the $b$ and $\tau$ sector are set to zero. The
different plots show the variation with $\tb$, $\tb = 2, 5, 20$. In
the $\cp$-conserving case for the above chosen soft SUSY-breaking
parameters, $\tb = 2$ is already excluded by Higgs boson
searches~\cite{tbexcl}. However, in the $\cp$-violating case this limit is
weakened~\cite{mhiggsCPXRG2} 
due to possible suppressed production cross section and/or suppressed
decays of the lightest Higgs boson to $b$ quarks, see
\refse{subsec:numanalhiggscouplings}. In each plot different values
for the charged Higgs boson masses have been chosen, 
$\MHp = 150, 200, 300, 500 \gev$. The largest effects of the phases
are observed for small $\tb$ and small $\MHp$. For large $\MHp$ the
effects of the $\cp$-violating phases become negligible small.

A numerical comparison with e.g. Fig.~3  in \citere{mhiggsCPXRG1}
shows agreement better than 10\% for not too large phases, 
$\phi_{\At} \lsim 0.8$. A larger phase corresponds to larger mixing in
the $\Stop$~sector. This, on the one hand, makes the corrections and
thus the uncertainties in the Higgs sector larger. On the other hand,
the RG improved EP calculations tends to loose accuracy for too large
mixing in the $\Stop$~sector.
The agreement improves slightly if a comparison with the more
complete result of \citere{mhiggsCPXRG2} (see e.g.\ Fig.~1) 
is performed. Furthermore, 
it has been shown in \citere{bse} that differences in the Higgs
boson masses arising from different renormalizations can be
significant, especially for large $\Stop$~mixing. 
Therefore 
agreement better than 5-10\% cannot yet be expected for all parameter
sets due to the different renormalizations employed and the yet more
complete evaluation performed in the RG improved 
EP calculation.

\begin{figure}[htb!]
\vspace{5em}
\begin{center}
\mbox{
\epsfig{figure=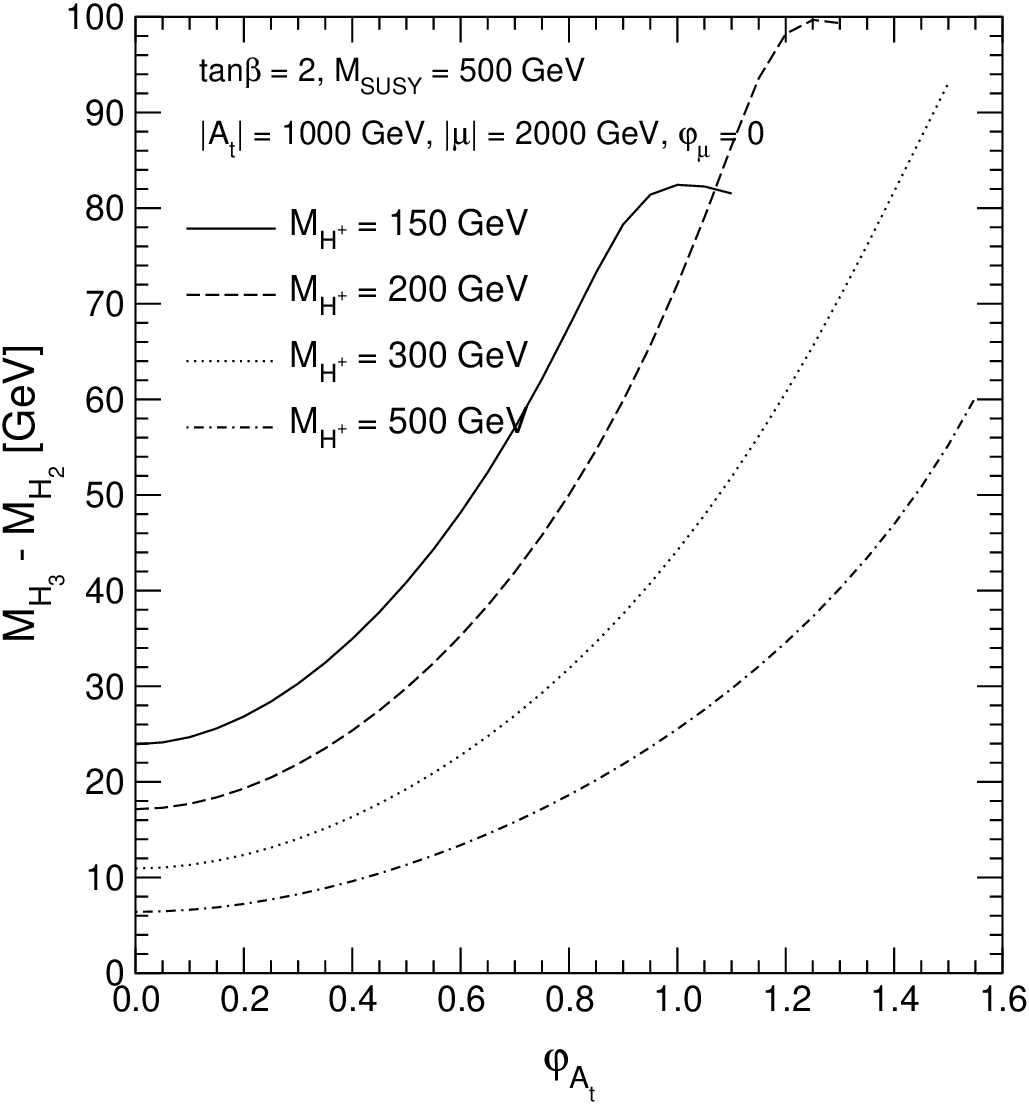,width=7cm,height=7.0cm} 
\hspace{1em}
\epsfig{figure=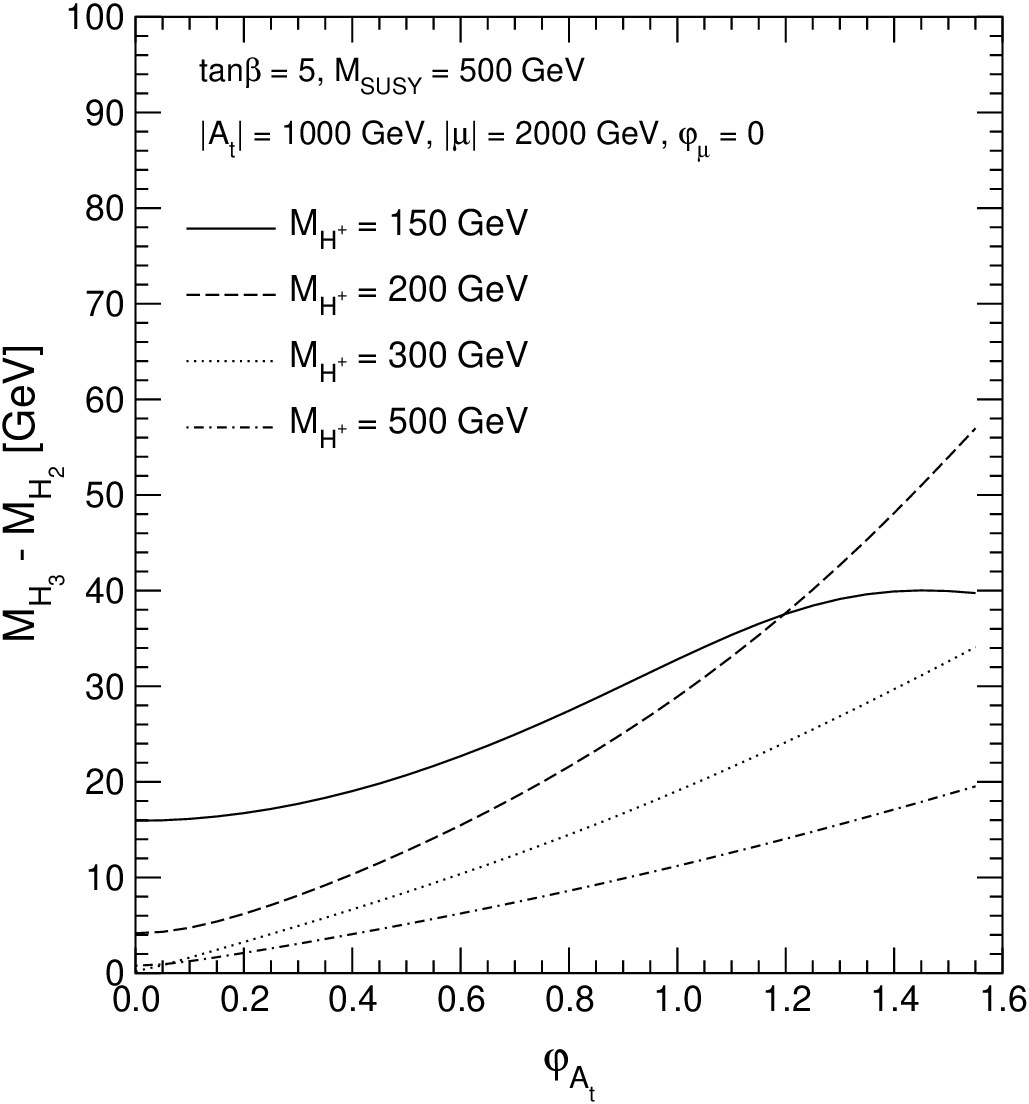,width=7cm,height=7.0cm}}

\vspace{3em}
\mbox{
\epsfig{figure=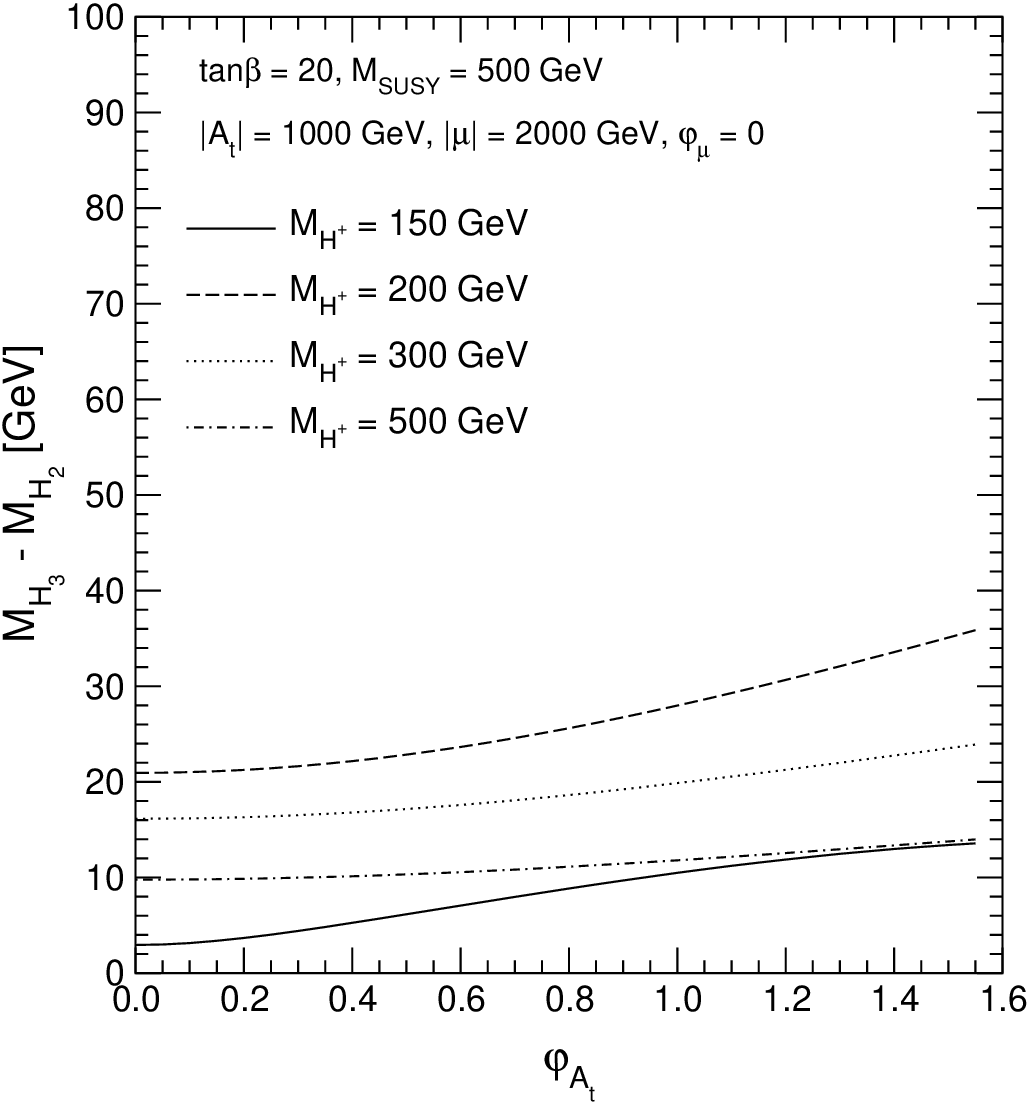,width=7cm,height=7.0cm} 
\hspace{1em}
\epsfig{figure=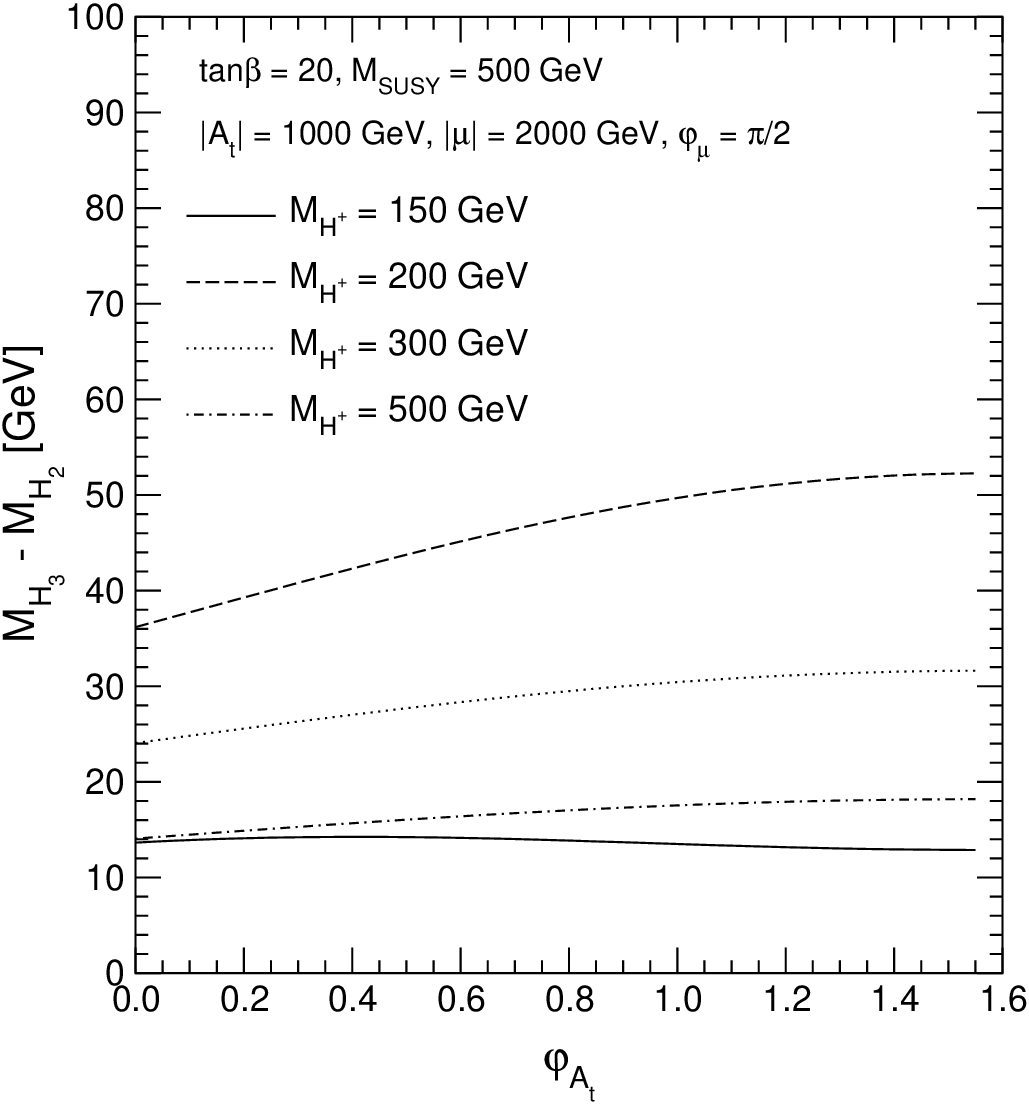,width=7cm,height=7.0cm}}
\end{center}
\vspace{1em}
\caption[]{
The mass difference between the two heavy Higgs boson, $\mHd - \mHz$,
is shown as a function of
$\phiat$ for different values of $\MHp$. 
In the first three plots $\phimu$ is set to zero and $\tb$ is chosen
as $2, 5, 20$. In the last plot $\phimu = \pi/2$ and $\tb = 20$ is
taken. The other parameters are $\msusy = 500 \gev, |\At| = 1000 \gev$
and $|\mu| = 2000 \gev$.
}
\label{fig:mh3-mh2_phiat}
\vspace{1em}
\end{figure}

In \reffi{fig:mh3-mh2_phiat} the mass difference of the two heavier
Higgs bosons, $\mHd - \mHz$, is shown as a function of $\phiat$. The
other parameters are chosen as in \reffi{fig:mh1mh2_phiat}. 
A large enhancement of the mass difference can be observed for small
$\tb$. The agreement with \citeres{mhiggsCPXRG1,mhiggsCPXRG2} 
is found at the same
level as for \reffi{fig:mh1mh2_phiat}.

\subsection{Higgs boson couplings}
\label{subsec:numanalhiggscouplings}

\begin{figure}[htb!]
\vspace{5em}
\begin{center}
\mbox{
\epsfig{figure=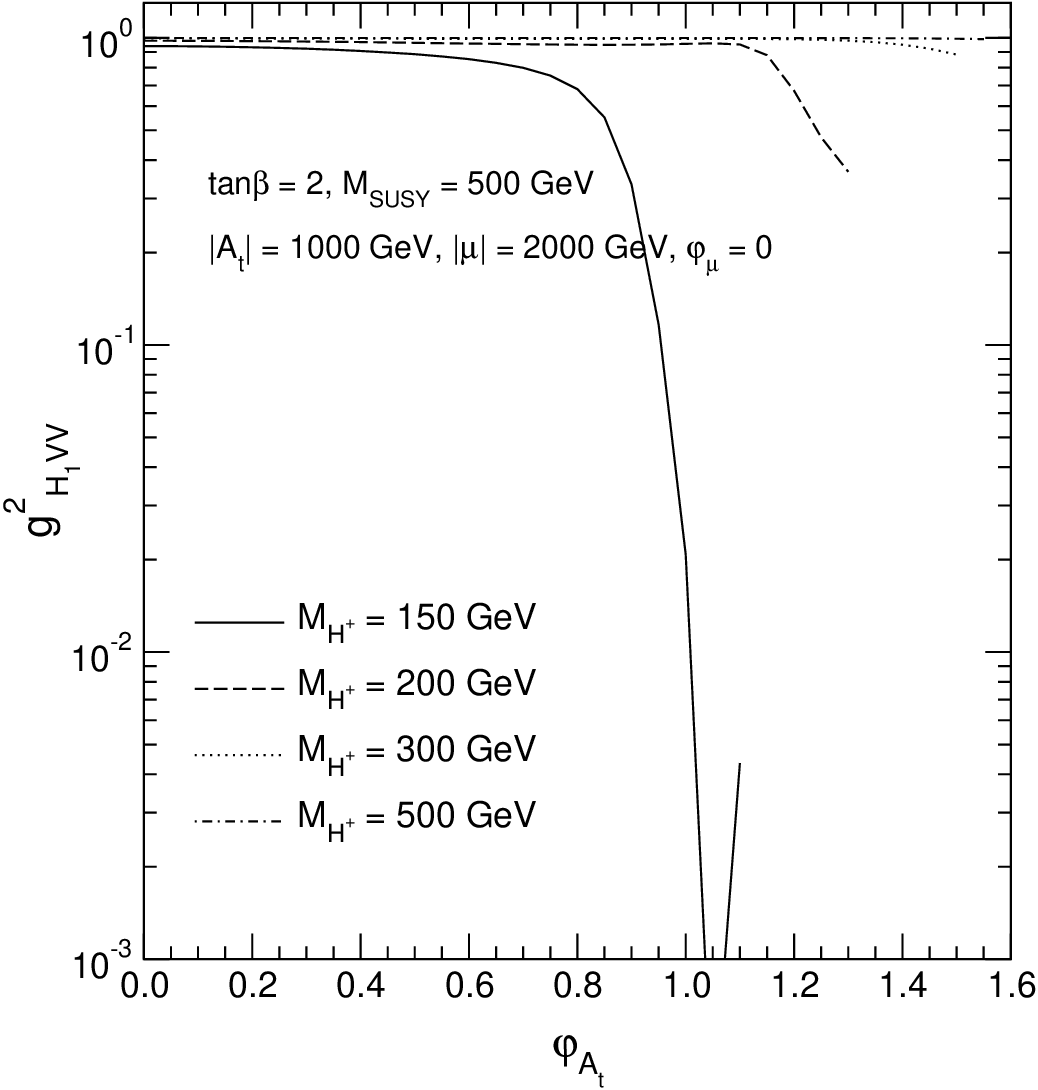,width=7cm,height=7.0cm} 
\hspace{1em}
\epsfig{figure=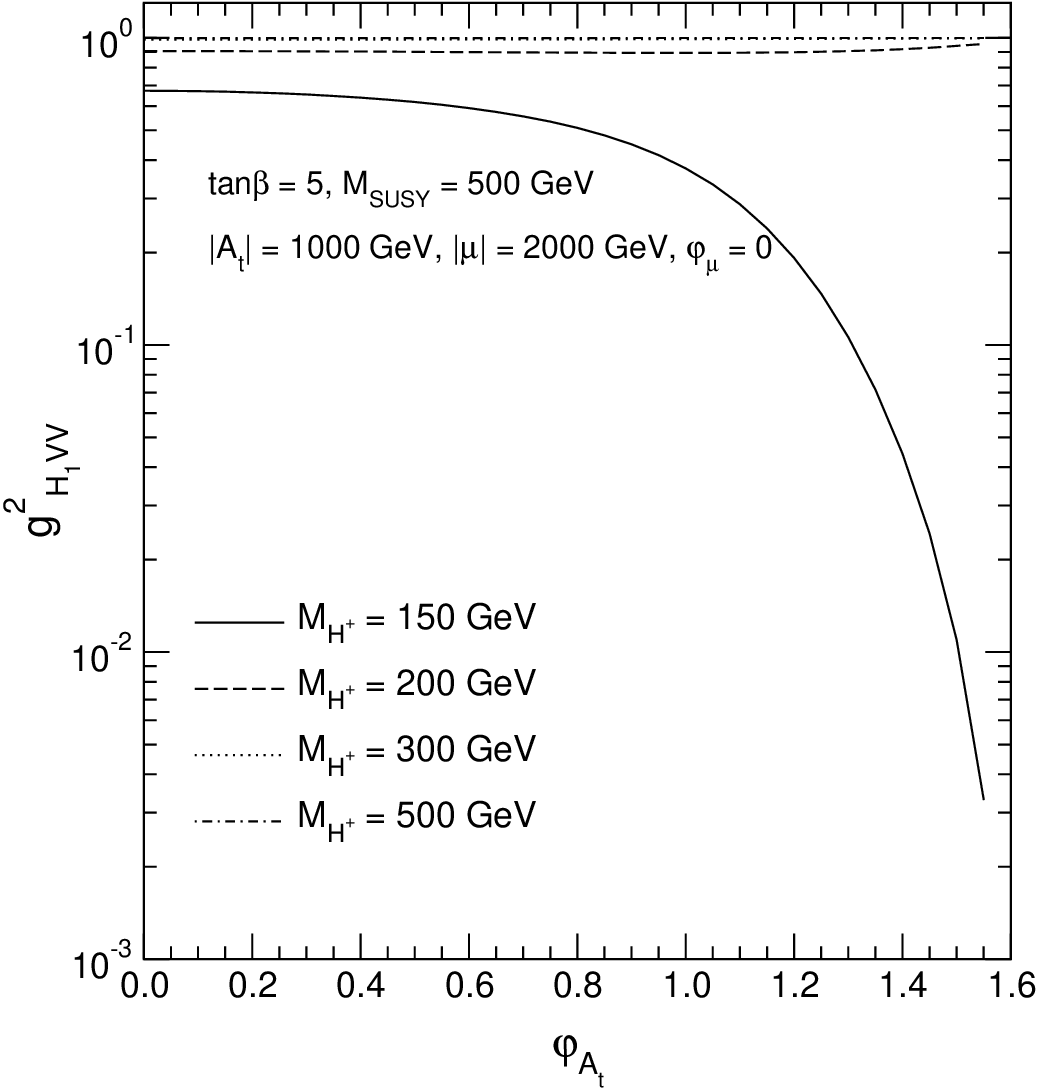,width=7cm,height=7.0cm}}

\vspace{3em}
\mbox{
\epsfig{figure=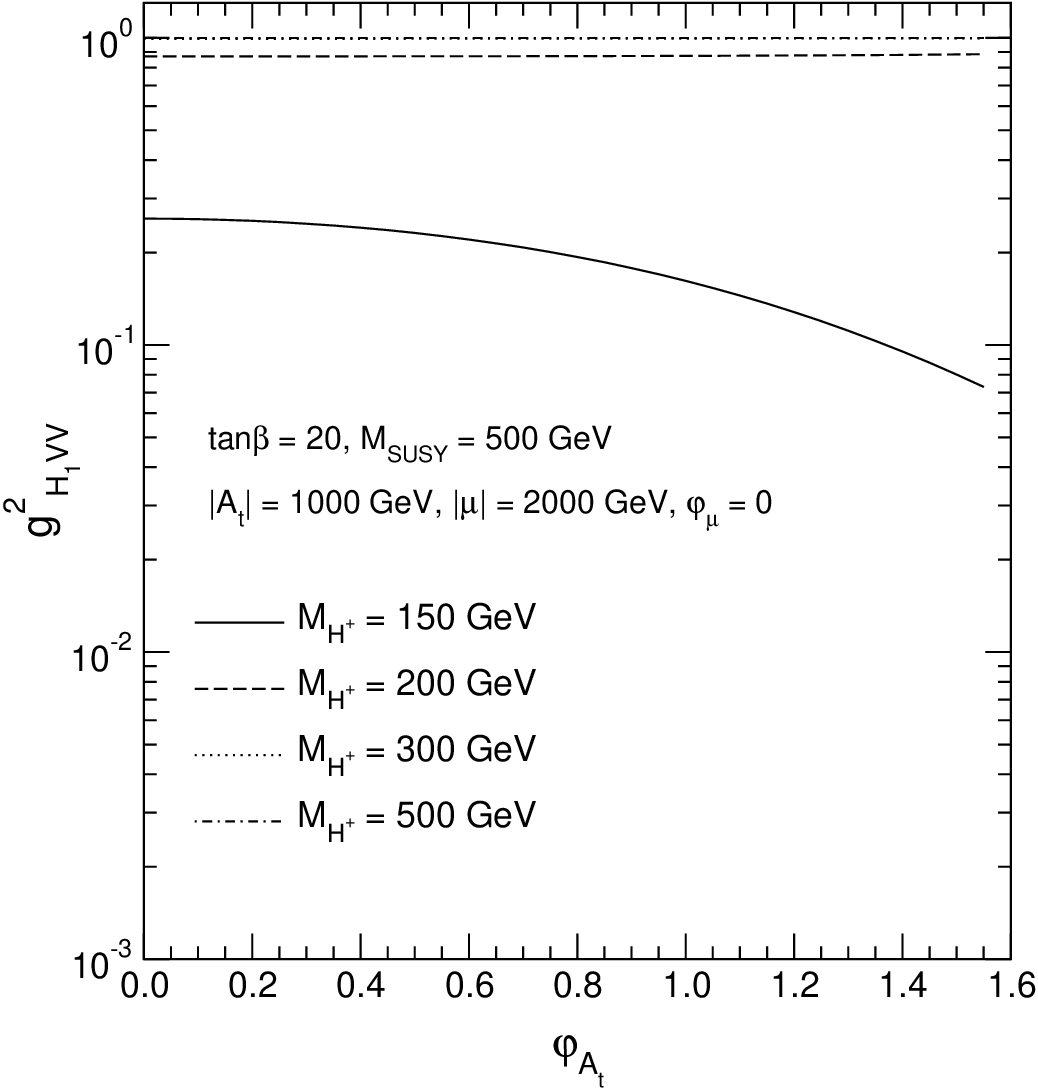,width=7cm,height=7.0cm} 
\hspace{1em}
\epsfig{figure=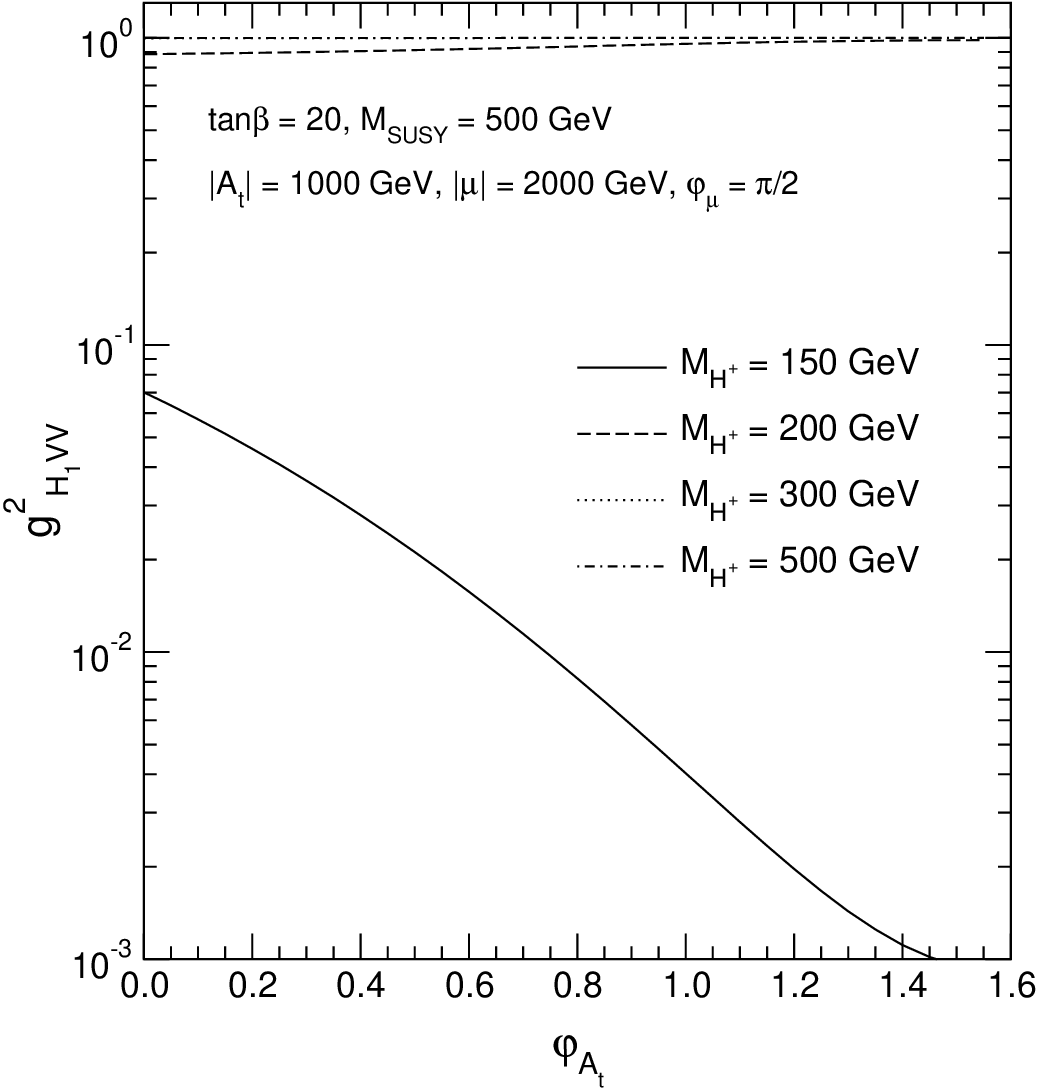,width=7cm,height=7.0cm}}
\end{center}
\vspace{1em}
\caption[]{
The coupling of the lightest Higgs boson to two gauge bosons (relative to its
SM value) is shown as a function of
$\phiat$ for different values of $\MHp$. 
In the first three plots $\phimu$ is set to zero and $\tb$ is chosen
as $2, 5, 20$. In the last plot $\phimu = \pi/2$ and $\tb = 20$ is
taken. The other parameters are $\msusy = 500 \gev, |\At| = 1000 \gev$
and $|\mu| = 2000 \gev$.
}
\label{fig:gH1VV_phiat}
\end{figure}

\begin{figure}[htb!]
\vspace{5em}
\begin{center}
\mbox{
\epsfig{figure=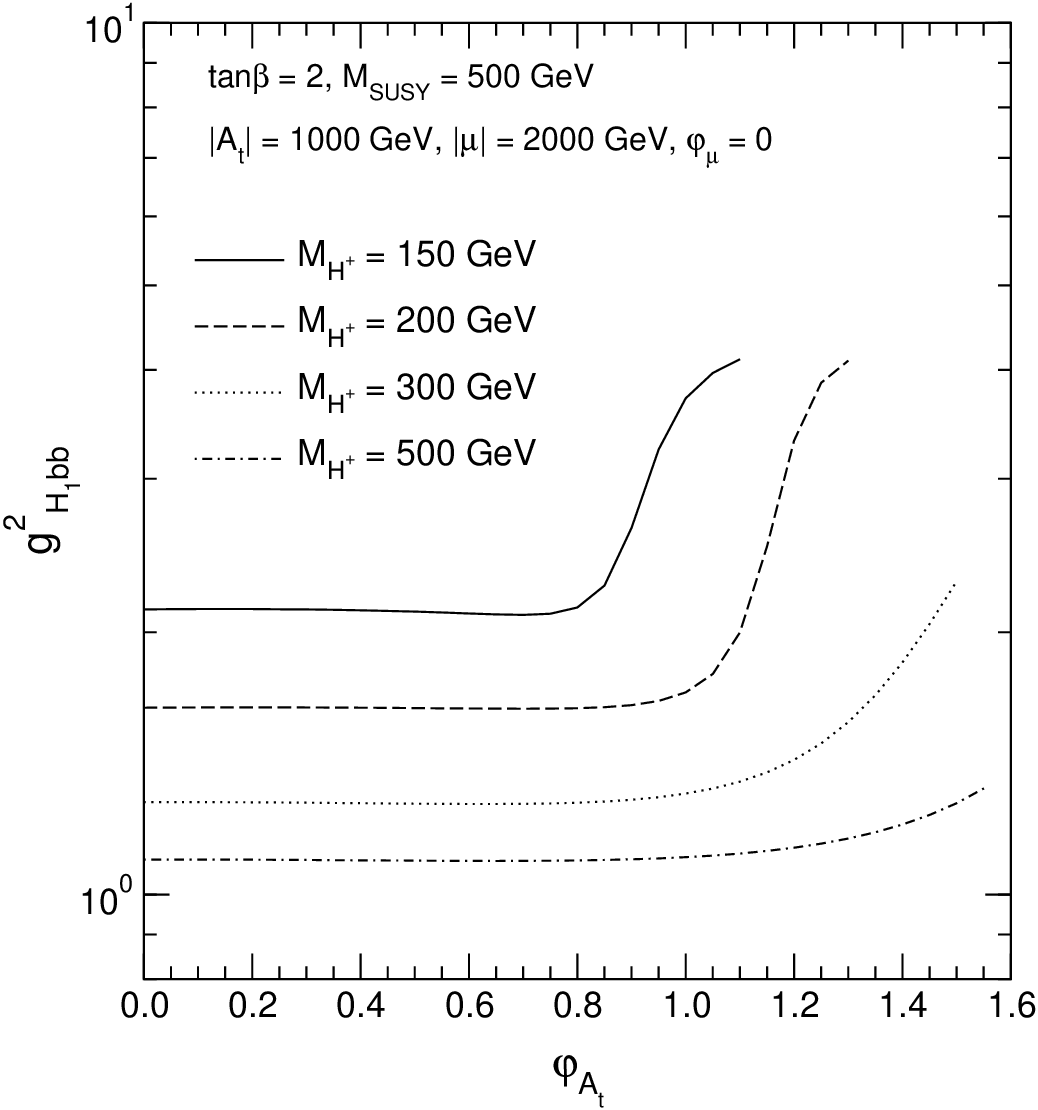,width=7cm,height=7.0cm} 
\hspace{1em}
\epsfig{figure=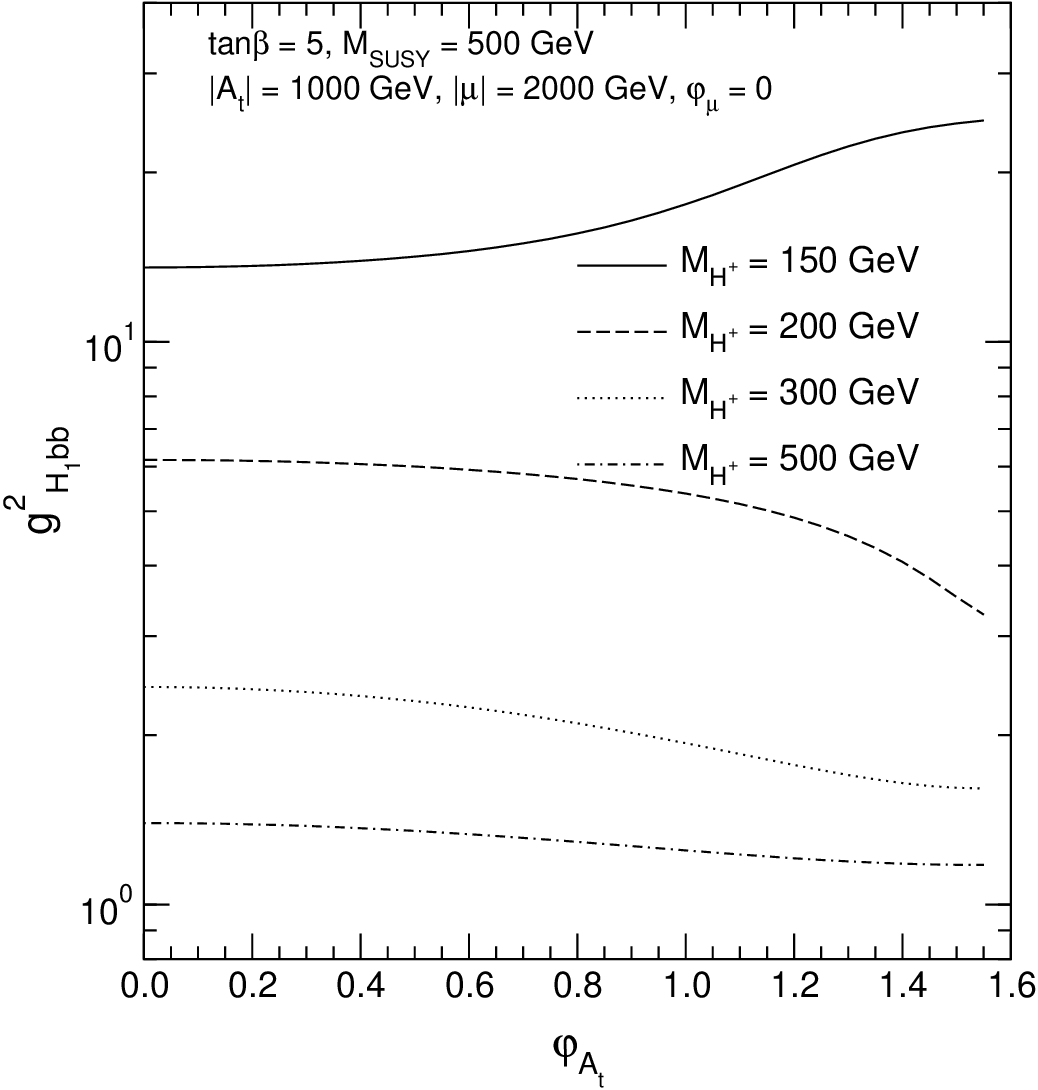,width=7cm,height=7.0cm}}

\vspace{3em}
\mbox{
\epsfig{figure=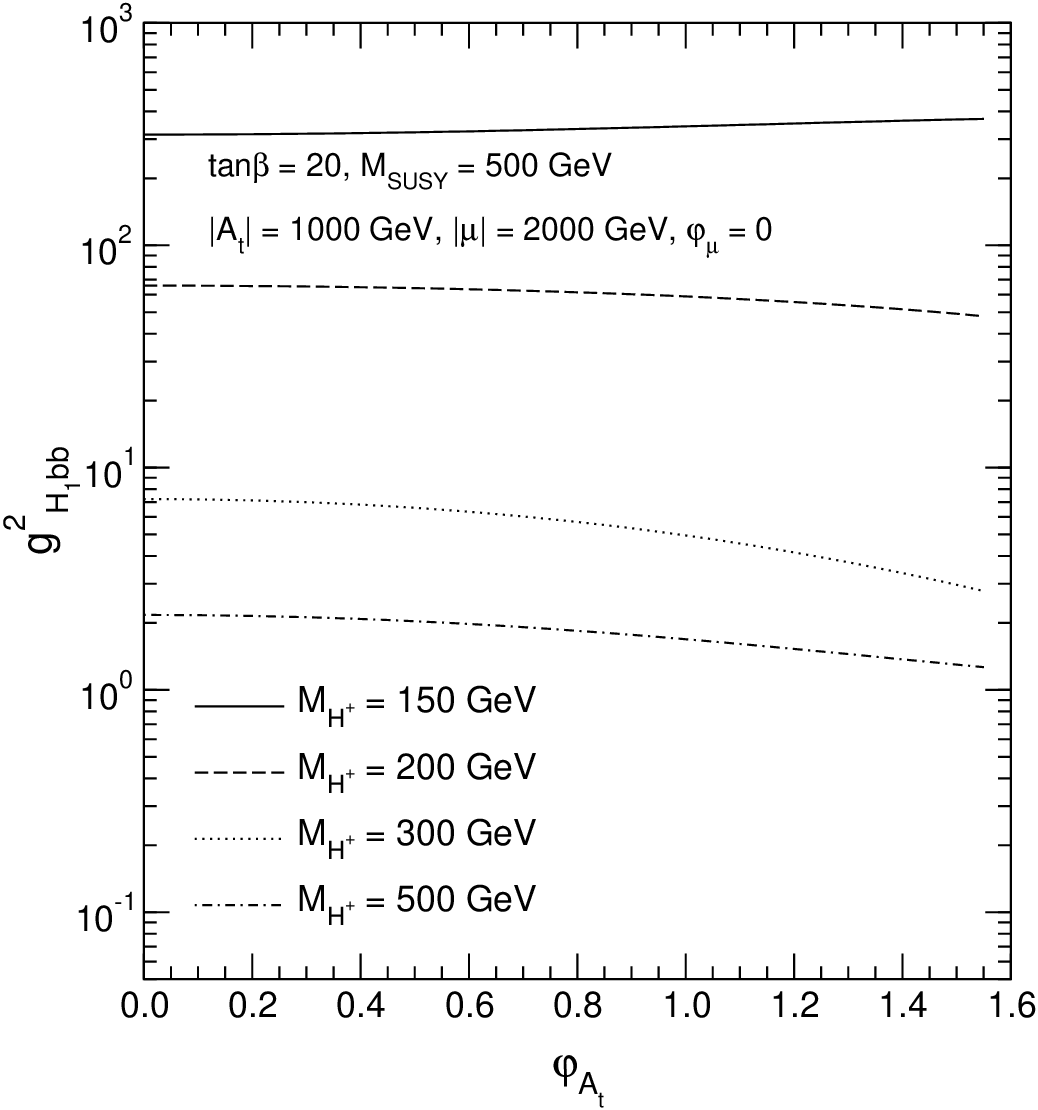,width=7cm,height=7.0cm} 
\hspace{1em}
\epsfig{figure=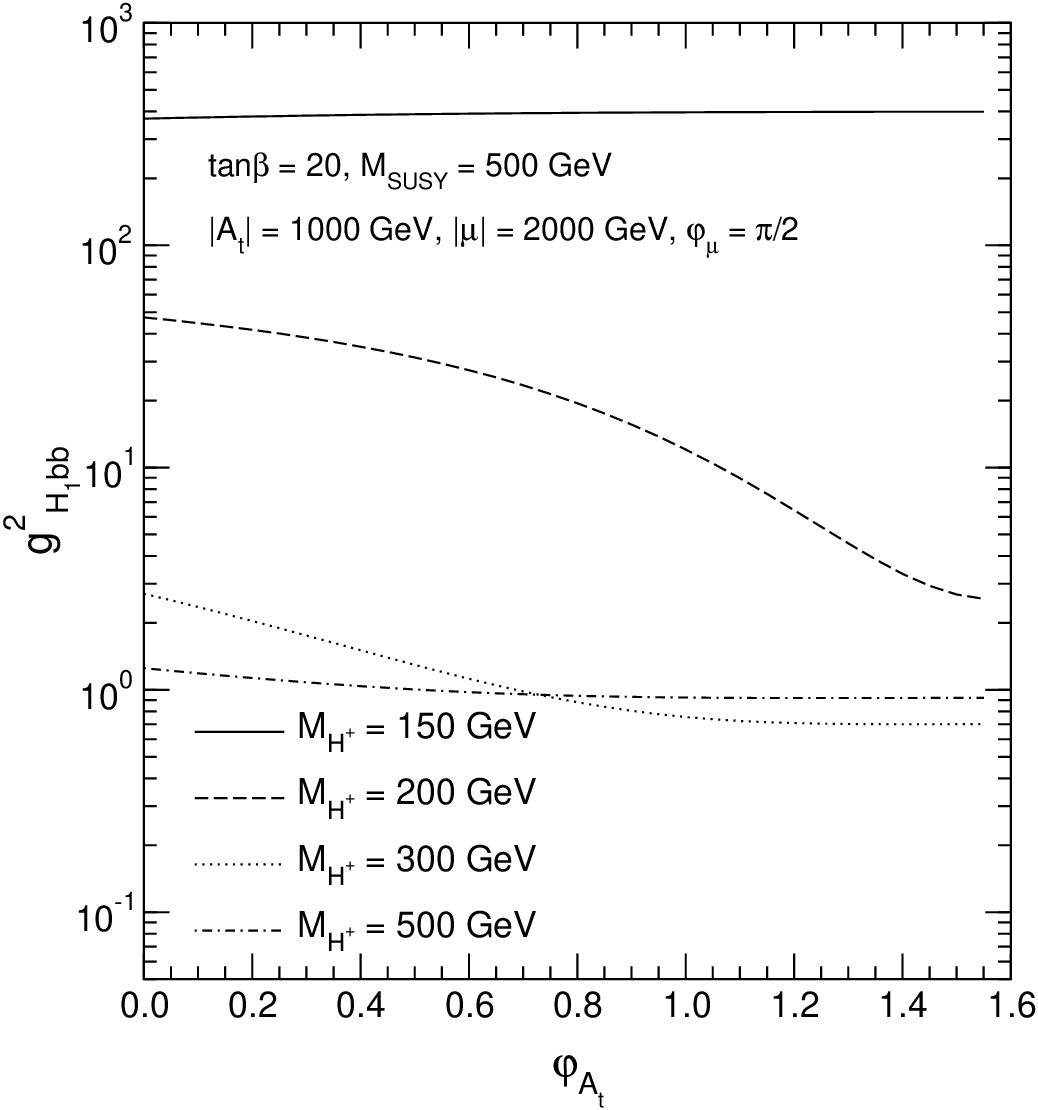,width=7cm,height=7.0cm}}
\end{center}
\vspace{1em}
\caption[]{
The decay rate of the lightest Higgs boson to $b$ quarks, 
$\Ga(H_1 \to b\bar b)$, relative to its SM value, is shown as a function of
$\phiat$ for different values of $\MHp$. 
In the first three plots $\phimu$ is set to zero and $\tb$ is chosen
as $2, 5, 20$. In the last plot $\phimu = \pi/2$ and $\tb = 20$ is
taken. The other parameters are $\msusy = 500 \gev, |\At| = 1000 \gev$
and $|\mu| = 2000 \gev$.
}
\label{fig:gH1bb_phiat}
\end{figure}

In \reffi{fig:gH1VV_phiat} the coupling of the lightest Higgs boson to
two SM gauge bosons, relative to its SM value, is shown as a function
$\phiat$. The other parameters are chosen as in \reffi{fig:mh1mh2_phiat}. 
Large suppressions occur for small values of $\MHp$. 
For $\MHp \gsim 250 \gev$ no suppression could be observed. For small
$\tb$ the suppression can amount several orders of magnitude, whereas
for large $\tb$ a suppression by a factor of 10 can be observed.
These results can be compared with the RG improved EP
approach, \citeres{mhiggsCPXRG1} Fig.~5 and \citere{mhiggsCPXRG2} Fig.~1. 
As for the Higgs boson masses, we find
reasonable agreement for not too large values of $\phi_{\At}$.

In \reffi{fig:gH1bb_phiat} the decay rate of the lightest Higgs boson to
$b$ quarks, $\Ga(H_1 \to b \bar b)$, relative to its SM value, 
is shown as a function $\phiat$. 
The other parameters are chosen as in \reffi{fig:mh1mh2_phiat}. 
The MSSM decay rate, although dependent on the complex phases, is
considerably larger than the SM one for most parts of the parameter
space. This renders the $b \bar b$ channel the main decay channel also
in the cMSSM.



\section{The Fortran code \fhfc}
\label{sec:feynhiggsfastc}

The results presented \refse{subsec:mt4eval} and
\refse{subsec:mhiggs2lcorr} are incorporated into the Fortran code
\fhfc . They are supplemented by the subleading \onel\
corrections from 
the $t/\Stop$~sector~\cite{mhiggslle} as well as by the full
logarithmic \onel\ 
corrections from all other sectors of the MSSM, obtained in the RG
approximation~\cite{mhiggsRG}.  

In the front-end of the code, the user can specify the
input parameters, including all relevant complex phases. 
This part can be manipulated at the user's will.
The main part of the code consists of the routines needed for the
evaluation of the higher-order corrections to the neutral Higgs boson
mass matrix, and should not be manipulated.

\noindent
\fhfc\ evaluates the following items in the cMSSM Higgs sector:
\begin{itemize}
\item
the three neutral Higgs boson masses
\item
the effective couplings of one neutral Higgs boson to two SM gauge
bosons and of two neutral Higgs bosons to a $Z$~boson
\item
the changes in the branching ratio 
for a Higgs decaying to SM fermions
\end{itemize}

\noindent
Furthermore the following ``check items'' are evaluated:
\begin{itemize}
\item
the SUSY corrections to the $\rho$-parameter, 
coming from the $\Stop/\Sbot$ sector.
(The complex phases enter only via their effective change of
the $\Stop$~and $\Sbot$~masses, where they can enlarge the splitting
and increase the contribution to $\De\rho$.) The SUSY corrections are
implemented in $\oa$ and $\oaas$, where the gluino-exchange
corrections, which go to zero for large $\mgl$ have been
omitted~\cite{delrhosusy2loop}. 
A value of $\De\rho$ outside the experimentally preferred region of 
$\dr^{\SU} \simleq 3\times 10^{-3}$~\cite{pdg} indicates experimentally
disfavored $\Stop$  and $\Sbot$ masses.
\item
the EDM of the electron and
the neutron, following the calculation of \citere{edm}%
\footnote{
We thank C.~Schappacher for providing the corresponding Fortran code.
}
~with the
convention of common soft SUSY-breaking parameters for up- and
down-type squarks. Values outside the experimentally allowed ranges
indicate either too large $\cp$-violating phases or demand heavier
squarks in the first two families~\cite{heavy12}.
\end{itemize}

\noindent
The code can be obtained from the \fh~\cite{feynhiggs} home page:
{\tt www.feynhiggs.de} .



\section{Conclusions}

We have presented the application of the Feynman-diagrammatic method
and the on-shell renormalization scheme to radiative corrections in
the Higgs sector of the MSSM with complex phases.
This provides a 
complementary method to the (renormalization group improved) Effective
Potential approach that has been used so far for phenomenological
analyses. 
The presented set-up can then be used for a detailed study of
the cMSSM Higgs sector in the FD/on-shell approach.

The general FD/on-shell method has been analyzed. Details about the
renormalization in the on-shell scheme and the derivation of the
renormalized Higgs boson self-energies have been presented. As an
example the leading fermionic corrections to the cMSSM Higgs sector
have been calculated analytically, making use of the recently
completed MSSM model file for \fa~3. After showing the generic
applicability of the approach, some numerical examples have been
calculated. The leading fermionic corrections have been supplemented
by the leading \twol\ corrections. Results have been obtained for the
masses of 
the neutral cMSSM Higgs bosons, their couplings to SM gauge bosons and
their couplings to SM fermions. 
Reasonable agreement better than 10\% with the RG improved EP method
has been 
found for not too large mixing in the scalar top sector. 

Finally the public Fortran code \fhfc\ has been presented. It provides
the evaluation of the masses and couplings of the cMSSM Higgs bosons
in dependence of the relevant cMSSM parameters, including all possible
complex phases. Besides the leading fermionic \onel\ and the leading
\twol\ corrections, also the full logarithmic \onel\ contributions,
taken over from the real MSSM, have been implemented. 
The code is obtainable at {\tt www.feynhiggs.de} .


\subsection*{Acknowledgements}
We thank S.~Dawson, M.~Drees, A.~Pilaftsis and C.~Schappacher 
for helpful
discussions and W.~Hollik and C.~Wagner for a critical reading of the
manuscript. 
We furthermore thank T.~Hahn, C.~Schappacher and other members of the TP,
Universit\"at Karlsruhe, Germany, for their effort put into \fa\ and
the new MSSM model file. 



\end{document}